\documentclass{acm_proc_article-sp}

\usepackage{balance}  
\usepackage{graphics} 
\usepackage{times}    
\usepackage[hyphens]{url}      
\usepackage{dsfont,enumitem}

\makeatletter
\def\url@leostyle{%
  \@ifundefined{selectfont}{\def\UrlFont{\sf}}{\def\UrlFont{\small\bf\ttfamily}}}
\makeatother
\def\pprw{8.5in}
\def\pprh{11in}

\usepackage[pdftex]{hyperref}
\hypersetup{
pdftitle={},
pdfauthor={LaTeX},
pdfkeywords={SIGCHI, proceedings, archival format},
bookmarksnumbered,
pdfstartview={FitH},
colorlinks,
citecolor=black,
filecolor=black,
linkcolor=black,
urlcolor=black,
breaklinks=true,
}

\usepackage{subfig}

\newcommand\Mark[1]{\textsuperscript#1}

\def\sharedaffiliation{
\end{tabular}
\begin{tabular}{c}
}
\newcommand{\putat}[3]{\begin{picture}(0,0)(0,0)\put(#1,#2){#3}\end{picture}}

\newcommand{\fig}[1]{Figure~\ref{#1}}
\newcommand{\furl}[1]{\footnote{\url{#1}}}

\newcommand{\lac}{App trend pattern}
\usepackage{rotating}
\newcommand{\andusers}{339,842}
\newcommand{\playapps}{213,667}
\newcommand{\cats}{47} 
\newcommand{\datalines}{13,779,666}

\begin{document}

\title{Sovereignty of the People: \\
There's more to Relevance than Downloads}


\numberofauthors{3}
\author{
 \alignauthor Stephan Sigg\\
   \affaddr{Aalto University}\\
   \affaddr{\tt stephan.sigg@aalto.fi}
 \alignauthor Eemil Lagerspetz\Mark{1}\\
 Ella Peltonen\Mark{1}
 \alignauthor Petteri Nurmi\Mark{1}\Mark{2}\\
 Sasu Tarkoma\Mark{1}\Mark{2}
\sharedaffiliation
   \affaddr{\Mark{1} Department of Computer Science, University of Helsinki}\\
   \affaddr{PO Box 64, FI-00014, University of Helsinki, Finland}\\
   \affaddr{\tt firstname.lastname@cs.helsinki.fi} \\
	\affaddr{\Mark{2} Helsinki Institute for Information Technology HIIT}\\
	\affaddr{Department of Computer Science, University of Helsinki}
}


\maketitle
\putat{305}{168}{\includegraphics[height=1cm]{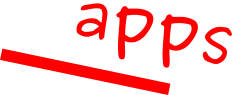}}
\begin{abstract}

The value of mobile apps is traditionally measured by metrics such as the number of downloads, installations, or user ratings. 
A problem with these measures is that they reflect actual usage at most indirectly. 
Indeed, analytic companies have suggested that retention rates, i.e., the number of days users continue to interact with an installed app are low. 
We conduct the first independent and large-scale study of retention rates and usage behavior trends in the wild. 
We study their impact on a large-scale database of app-usage data from a community of \andusers{} users and more than \playapps{} apps. 
Our analysis shows that, on average, applications lose 70\% of their users in the first week, while very popular applications (top 100) lose only 45\%. 
It also reveals, however, that many applications have more complex usage behavior patterns due to seasonality, marketing, or other factors. 
To capture such effects, we develop a novel app-usage behavior trend measure which provides instantaneous information about the "hotness" of an application. 
We identify typical trends in app popularity and classify applications into archetypes. 
From these, we can distinguish, for instance, trendsetters from copycat apps. 
In our results, roughly 40\% of all apps never gain more than a handful of users.
Less than 0.4\% of the remaining 60\% are constantly popular, 1\% flop after an initial steep rise, and 7\% continuously rise in popularity. 
We conclude by demonstrating that usage behavior trend information can be used to develop better mobile app recommendations.
With the proposed usage-based measures (retention and trend), we are able to shift sovereignty in app recommendations back to where it really matters: actual usage statistics, in contrast to download count and user ratings which are prone to manipulation by people.

\end{abstract}


\linebreak
\section{Introduction}
\label{intro}
How valuable and good is my mobile application? 
With the popularity of mobile apps continuing rapid growth, answering this question has become far from straightforward. 
According to numbers published in 2016\furl{http://www.appbrain.com/stats} \furl{http://appshopper.com/}, the most popular application stores, Google Play and Apple's App Store, both feature more than 1.6 million applications already. 
Given this plethora of apps, many simply vanish in this digital cornucopia. 
Indeed, studies on app marketplaces have shown that overall rating and the nature of user reviews are key drivers in application download decisions~\cite{Harman:2012,kim:2011}. 
This sovereignty of people and individuals reduces the chances of apps within a small community of users to ever become known within the greater public due to the small number of ratings, even if rating score is high and usage frequent. 
Ratings are well-known to be vulnerable to spam and rating fraud~\cite{Chandy:2012,Jindal:2008}. 
At the same time, studies on app usage behavior have shown that downloads and ratings can be misleading as users often install apps to try them out~\cite{Baeza-Yates:2015,Bohmer:2013} and rate apps negatively for a wide range of reasons, such as technical problems or lack of features~\cite{khalid15what}. 


Instead of relying on installation counts or ratings, analytics companies often use {\em retention rate} to measure the success of mobile apps\furl{http://info.localytics.com/blog/the-8-mobile-app-metrics-that-matter}. Retention rate reflects the number of days that users continue to interact with an application after installing it. 
Among others, retention rate can inform developers of the engagement with their app or it can be used by advertisers to determine the price of advertising. 
Reports by analytics companies\furl{http://andrewchen.co/new-data-shows-why-losing-80-of-your-mobile-users-is-normal-and-that-the-best-apps-do-much-better/} suggest that retention rates of most applications are low, since only a small fraction of users continues to use the apps. 
Even $80\%$ drop of usage is reported common within a week from the first use. 
However, thus far no independent information about retention rates of mobile apps in the wild has been available. 

As our \textbf{first contribution}, we present results from the first independent study of retention rates in the wild. We perform our study on usage behavior data recorded in the frame of the Carat project~\cite{caratSensys2013} collected from \andusers{} Android devices over a period of three years (June 2012 -- July 2015). 
Our results indeed confirm that, on average, applications lose $70\%$ of their users within a week from first use. 
However, these effects are mediated by the number of users an application has. 
For applications with $10$ or more users, retention rates below $30\%$ are rare, and retention rates increase with the popularity of the app. 
We also demonstrate that retention rates alone are not a sufficient measure of an app's success as they ignore fluctuation in instantaneous usage, the effects of seasonality, external factors, and other long-term usage behavior trends. 
To capture and quantify these effects, novel metrics for mobile analytics are required.

\begin{figure}
	\includegraphics[width=\columnwidth]{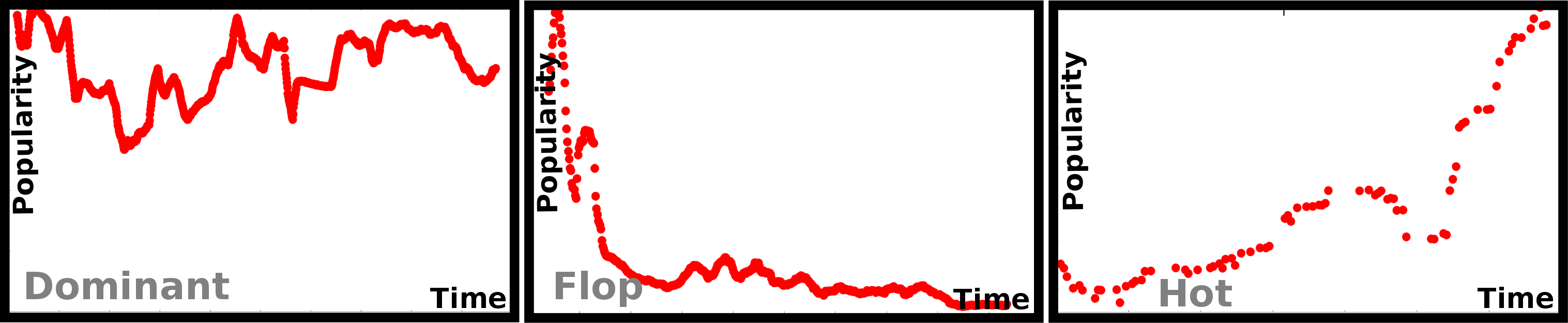}
	\caption{{\sl Flops} quickly lose in usage after a high peak, 
	{\sl Hot} apps continuously rise in popularity,
	{\sl Dominant} apps feature stable usage and {\sl Marginal} apps never reach a significant user base.
}
	\label{figureLifecycles}
\end{figure}

As our \textbf{second contribution}, we develop a novel app-usage trend filter, which provides instantaneous information about the "hotness" of an application. 
Our filter captures the relative popularity of apps based on daily use. 
It indicates behavior trends regardless of absolute volume and categorizes apps to \textit{\lac{}s} that can be used to predict the future relevance of an app, or to categorize apps. 
Characteristic trend patterns include {\sl Flop}, {\sl Hot}, {\sl Dominant} or {\sl Marginal apps} (\fig{figureLifecycles}).
We validate the benefits of our app-usage filter through a large-scale analysis of mobile app usage trends in the wild. 
We first consider carefully selected applications, which are widely acknowledged as hits or flops, and demonstrate that the usage behavior trend-filter can correctly identify them. 
We then apply our filter to all apps in the dataset, characterizing their popularity. 
Our results show 40\% are {\em Marginal} apps, and in the remaining 60\%, only 0.4\% are {\em Dominant}, 1\% are {\em Flops}, and 7\% are {\em Hot}. 

As a \textbf{practical use case}, we analyse the performance of a state-of-the-art mobile app recommender~\cite{appjoy-2011} with respect to trends. 
Our analysis shows that only $3.6\%$ of the recommendations are for apps which are currently rising in popularity, and that overall recommendations have low novelty and temporal diversity. 
We also demonstrate that the accuracy of the recommendations can be improved by considering trend information. 
Trend information not only helps the app developers, but can also help users to discover more relevant apps.

Summarising, our usage-based measures retention and trend are able to shift sovereignty in app recommendations back to where it really matters: actual usage statistics, in contrast to download count and user ratings which are prone to manipulation by people.

\section{Related Work}
Mobile app recommendation systems utilise a multitude of features to rate the relevance or quality of a respective app. 
Among these, user reviews are a prominent source for app recommendation systems~\cite{Fu13why}. 
However, empirical studies have shown that reviews typically contain several topics, which are seldom reflected by the overall rating~\cite{pagano13user,khalid15what}. 
Motivated by these studies, several works on using sentiment analysis and summarisation techniques for mining app reviews have been proposed. 
Chen et al.~\cite{chen14arminer} identify reviews that are most informative to the developers, whereas Guzman and Maalej~\cite{guzman14how} use sentiment analysis and language processing to extract user opinions for different features in a mobile app. 
Recommendations, however, are prone to fraud and are inaccurate and noisy as they are based on free-form textual descriptions.
Guerrouj et al.~\cite{guerrouj15influence} mine feature requests from StackOverflow discussions and demonstrate that the amount of feature requests has a significant effect on overall ratings, with popular and highly rated applications receiving significantly fewer changes and feature requests than less successful ones.
Jovian et al. point out that version information should further impact the recommendation score as the quality of an app might be affected by the change in version~\cite{lin2014new}.
This is implicitly also covered by our approach as changes in the usage trend might result from a version update. 

In order to improve the above mentioned global solutions in which recommendations are identical for all users, individual preferences are considered.
Peifeng et al. argue that an app recommendation system has to take into account also the set of already installed apps~\cite{yin2013app}.
They compare a 'tempting' value of a new application to a 'satisfactory' value of already installed applications of the same type. 
A recommendation is then provided only if the former exceeds the latter.
Another example is AppJoy~\cite{appjoy-2011}, which employs item-based collaborative filtering to recommend apps based on personalized usage patterns. 
AppBrain\furl{http://www.appbrain.com} compiles recommendations within the same category by monitoring the installation history of apps. 
Also, AppAware~\cite{Girardello-mobilechi-2010} provides recommendations by integrating the context information of mobile devices. 
Moreover, users often use several apps within the same category~\cite{zhong13google} and the overall usage session times tend to be short, and depend on a wide range of contextual factors~\cite{Bohmer-mobilechi-2011,reuver12smartphone}.
Responding to this observation, the AppTrends approach was proposed to consider actual usage data and to base the recommendation on frequency of co-usage of apps~\cite{AppRecommendation_Bae_2015}. 
Combinations of these individual recommendation systems to form multi-objective app recommendations have the potential to further improve accuracy in the recommendations~\cite{xia2014multi}.

All of the above solutions focus on the user but not on the actual popularity or quality of an app. 
Also, Petsas et al.~\cite{Petsas-imc-2013} demonstrated that user preferences tend to be highly clustered and following various trends over time, with users showing interest in a small set of app categories at a time. 
Our work complements existing solutions by providing mechanisms for analysing and understanding application usage relative to the dynamics of the app's instantaneous popularity in a marketplace. 

Some commercial app analytics tools, such as Google Mobile Analytics\furl{http://www.google.com/analytics/mobile} and Countly's Mobile Analytics\furl{http://www.count.ly}, follow a similar path by focusing on monitoring statistics of individual apps, covering information about usage session frequencies, lengths of usage sessions, extent of in-app purchases, and so forth. 
In contrast to our work, these solutions provide little insight into the usage of the app relative to other apps. 

Commercial trend mining systems include Google Trends\furl{google.com/trends}, which monitors the frequency of words in search queries related to real-world events, and the trending topics list of Twitter, which uses the frequency of hashtags and noun expressions to determine popular topics. 
Related academic works include detecting emerging trends in real-time from Twitter~\cite{benhardus13streaming,cataldi10emerging,mathioudakis10twitter}, mining of news discussions or other text documents for trends~\cite{Streibel-2011,popescul-2000}, and analysis of web behaviour dynamics~\cite{radinsky12modeling}. 
Our work is capable of operating solely on app usage information whereas these works operate on co-frequency patterns between words or n-grams.  
Moreover, these works only focus on a short time span, whereas we consider the entire lifecycle of an application. 

\section{Dataset}
\label{sec:dataset}

Our work considers data recorded with Carat~\cite{caratSensys2013}\furl{carat.cs.helsinki.fi}, a stock Android and IOS mobile app to propose personalized recommendations to improve a device's battery life. 
Carat uses energy-efficient and non-invasive instrumentation to record the state of the device, including the process list, and active apps. 
Carat has been deployed on over $800,000$ smartphones, roughly half of which are Android devices. 
The community of users that contribute data to Carat is spread all over the world, with users in roughly $200$ countries, and a strong presence in USA, most of Europe, India, and Japan. 

For this paper, we consider a subset of the data that contains measurements from \andusers{} Android devices over a period of three years (June 2nd, 2012 -- July 14th, 2015). 
We limit our analysis to Android devices as the data obtained on Android devices can be uniquely mapped into individual applications~\cite{truong14mobile}, whereas data obtained on iOS devices can not. 
As part of our analysis, we assess the popularity of apps within categories. 
To carry out this analysis, we combine the crowdsourced data with category information from Google Play. 
The resulting dataset includes a total of \datalines{} app usage records from \andusers{} users and \playapps{} apps in \cats{} categories.
We only use the leaf categories of Google Play. 
\textit{Games} and \textit{Family}, for instance, are groups of categories, while two of their individual categories are \textit{Games: Racing} and \textit{Family: Action}. 
The scale of the dataset we consider in our analysis is an order of magnitude larger than in previous works. 
For example, Harman et al.~\cite{Harman:2012} considered a dataset containing reviews from $30,000$ apps, whereas B{\"o}hmer et al.~\cite{Bohmer-mobilechi-2011} considered measurements from $4,125$ users and $22,626$ applications. 

{\em Ethical Considerations:} We analyse aggregate-level data which contains no personally identifiable information. 
The privacy protection mechanisms of Carat are detailed in~\cite{caratSensys2013}. 
Data collection by Carat is subject to the IRB process of University of California, Berkeley. 
Users of Carat are informed about the collected data and give their consent to use data from their devices.

\section{Analysis of Retention Rates}
The few existing academic studies on mobile app usage have mainly been able to characterize factors that drive download decisions~\cite{Harman:2012,kim:2011,Petsas-imc-2013}  without being able to determine what happens once the app has been installed on the device. 
While some studies have relied on measurements taken on the handsets, they have focused on overall usage and how that is influenced by contextual factors~\cite{Bohmer-mobilechi-2011,reuver12smartphone,ravindranath-osdi-2012}, leaving commercial reports by mobile analytics companies the only source of information about what happens once the app has been downloaded. 
Reports of analytics companies suggest that usage dwindles significantly after installation (i.e., low retention), and reportedly, even~$80\%$ of users stopping to use the application is common. 
In this section we present the first {\em independent} and large-scale study to investigate whether this indeed is the case.

\subsection{Retention Rate}
\label{sectionRetentionRate}
Retention rate on day $d$ is defined as the percentage of users that continue using the application $d$ days after first usage. 
To estimate retention rates, we identify for each user and application the first and last time the user launched the application. 
To ensure usage behaviour is correctly captured, for each app we only consider those users who had not been using the app within $7$ days of the last measurement day (July 14th 2015).

\begin{figure}
	\centering
	\subfloat[Retention rate largely depends on the initial number of users, with popular apps staying healthy for longer periods of time.]{\includegraphics[width=.48\columnwidth]{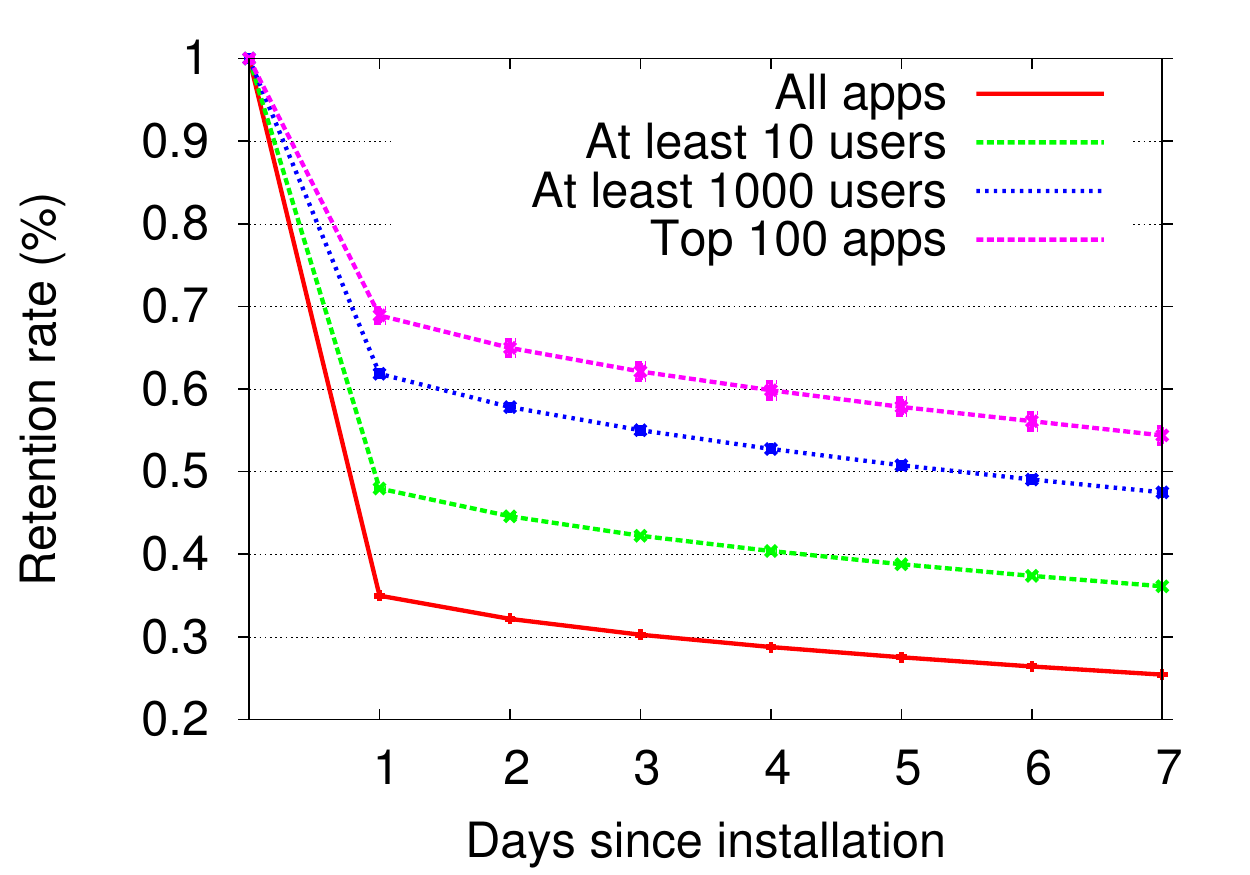}	\label{fig:retention-error}}\hfill
	\subfloat[Empirical cumulative distribution of retention rate for Apps with at least 1,000 users.]{\includegraphics[width=0.48\columnwidth]{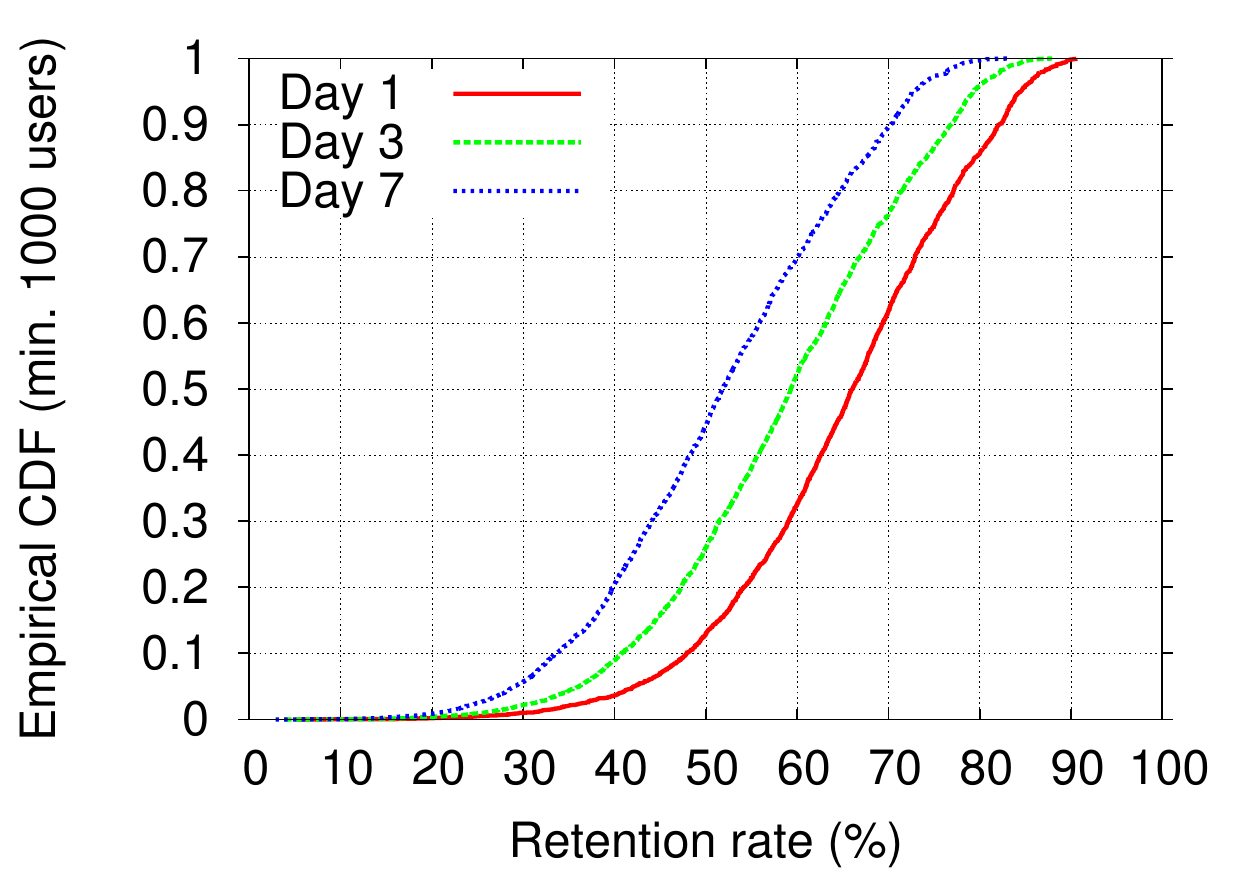}\label{fig:cdfplot-days-d137-10ALL}
	\label{fig:cdfplot-days-d137-1000}}
	\caption{Retention rates and their distribution}
\end{figure}

The retention rates of mobile apps in our dataset are illustrated in Figure~\ref{fig:retention-error}. From the figure we can observe that, while retention rates of many applications indeed are low, there are several apps with a healthier usage lifecycle. Overall, for all apps, the first day retention is as low as $36\%$ with $7$ day retention falling below $30\%$. However, our results also suggest that this effect is mainly due to many apps receiving only very few users. Indeed, retention rates for apps with at least $10$ users show much healthier behaviour, with first day retention being close to $50\%$. For apps with at least $1000$ users the same figure rises to $62\%$. For the most popular $100$ apps, first day retention is even as high as $68\%$ and after $7$ days the retention remains higher than $50\%$. In summary, our analysis supports the view that retention rates of apps tend to be low, with the usage witnessing a steep decline particularly after the first use. However, our analysis also calls into contention some of the claims made by analytics companies, indicating that the number of users mediates the retention of applications, with even apps with as few as $10$ users witnessing significantly better retention rates.

To further shed light on the retention patterns, Figure~\ref{fig:cdfplot-days-d137-1000} illustrates the empirical cumulative distribution function of retention rates for apps with at least $1,000$ users. 
In the plot we separately consider the retention rates of day $1$, day $3$ and $7$. 
These days were chosen for our results to be comparable with results published by analytics companies$^3$. 
From the plot we can see that high retention rates are rare. 
Indeed, only $10-40\%$ of apps have retention rates of $70\%$, and a mere $3\%$ of apps is able to achieve $75\%$ retention rate on day $7$. 
However, from the figure we also observe that extreme drops are rare, with less than $10\%$ of applications having retention rates below $30\%$, i.e., the $80\%$ drop reported in the literature is not common for apps that have been able to attract a sufficient user base. 

We observed similar patterns for apps with less than $1,000$ users. 
However, since there are orders of magnitude more apps with only a handful of users, as opposed to, for instance, hundreds of users, the retention rate of the entire data is then biased.
Apps with 100 users or less are very volatile in terms of retention rate, since a drop of a single user already decreases retention by 1\% or more. 
The plot for apps with at least 10 users follows a similar, but more jagged pattern, and rises much faster with retention rates around 10\% lower than in Figure~\ref{fig:cdfplot-days-d137-1000}.

This also further supports our earlier finding of retention being mediated by the size of the user base. 
To verify this, we used Spearman correlation to assess the statistical dependency between usage counts and retention rates. To limit potential biases and noise in the retention rate estimates, we only considered apps with at least $10$ users. The resulting analysis revealed the correlation to be statistically significant for all days ($d = 1, \rho = 0.199, p <.001; d = 3, \rho = 0.185, p <.001; d = 7, \rho = 0.165, p < 0.001$). For applications with higher usage count, correlations were slightly lower, but remained consistently significant.

For apps with only $10-15$ users, some fraction of the retention could be potentially explained by developers of the apps continuing to test and use their app. 
Unfortunately identifying these users from the Carat data is not possible. 

\subsection{Beyond Retention Rates}
\label{sec:dl}
\begin{figure*}
	\centering
	\includegraphics[width=0.975\linewidth]{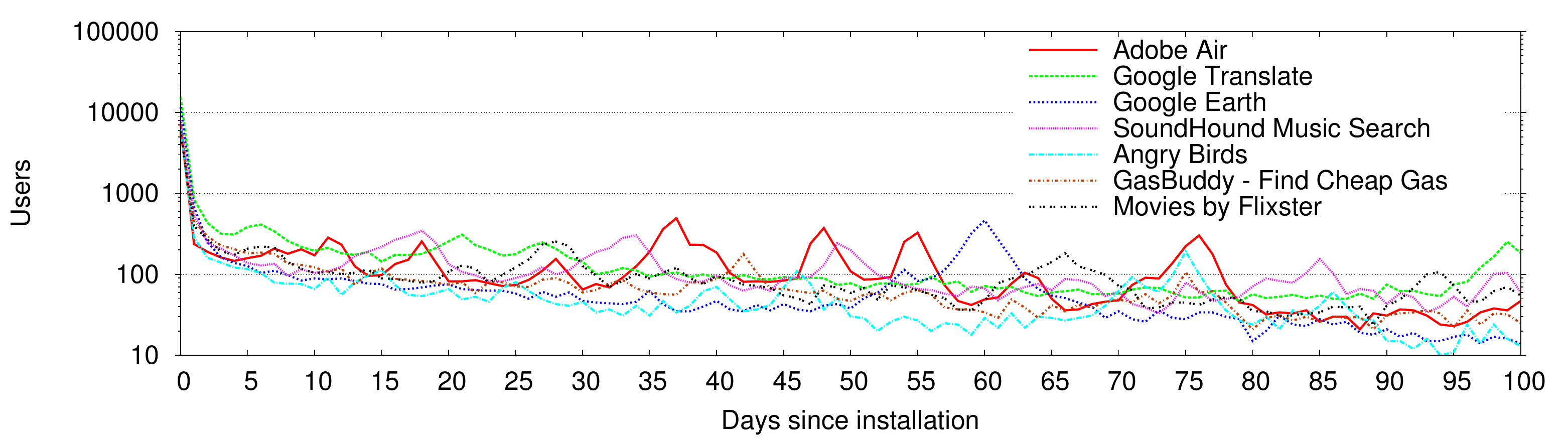}
	\caption{App usage patterns do not always follow a simple falloff graph as suggested by retention rate.}
	\label{fig:app-use-days}
\end{figure*}

While retention is able to reflect the long-term attractiveness of an app to individual users, it does not cover instantaneous popularity, trends or seasonal patterns. 
In particular, when the popularity of an app changes, this does not necessarily affect its retention rate. 
Furthermore, apps with seasonal usage patterns, such as recommendation of lunch places, nearby gas stations or vacation-related are unfairly treated by retention rates. 
This is in particular true when the time window is short, such as the often cited 1-day/3-day/7-day retention characteristics. 
For instance, Figure~\ref{fig:app-use-days} depicts usage patterns for exemplary applications from first day of usage up until $100$ days of usage. 
The selection of applications was done automatically using a peak detection algorithm that identifies significant peaks in usage after the initial slide in usage. 

In the figure, {\em GasBuddy} is a representative example of apps with clear seasonal patterns. 
It compares fuel prices at nearby gas stations. 
The use of the application dwindles after day $1$, but has recurrent spikes at biweekly and monthly intervals. 
Other examples include the \textit{Adobe Air} game store/platform (regular peaks), or \textit{Angry Birds} (many small peaks). 
Utilities, such as the music identifying \textit{SoundHound}, \textit{Google Translate}, and \textit{Google Earth} (both irregular peaks) are used on-demand as their value to users is situational.

To capture these effects and to provide a more accurate view of the usage of an application, we next develop a novel trend mining approach for capturing application lifecycles.

\section{Mining real-time trends}
\label{sec:trends}
\begin{figure*}[!t]
	\centering
  \includegraphics[width=\textwidth]{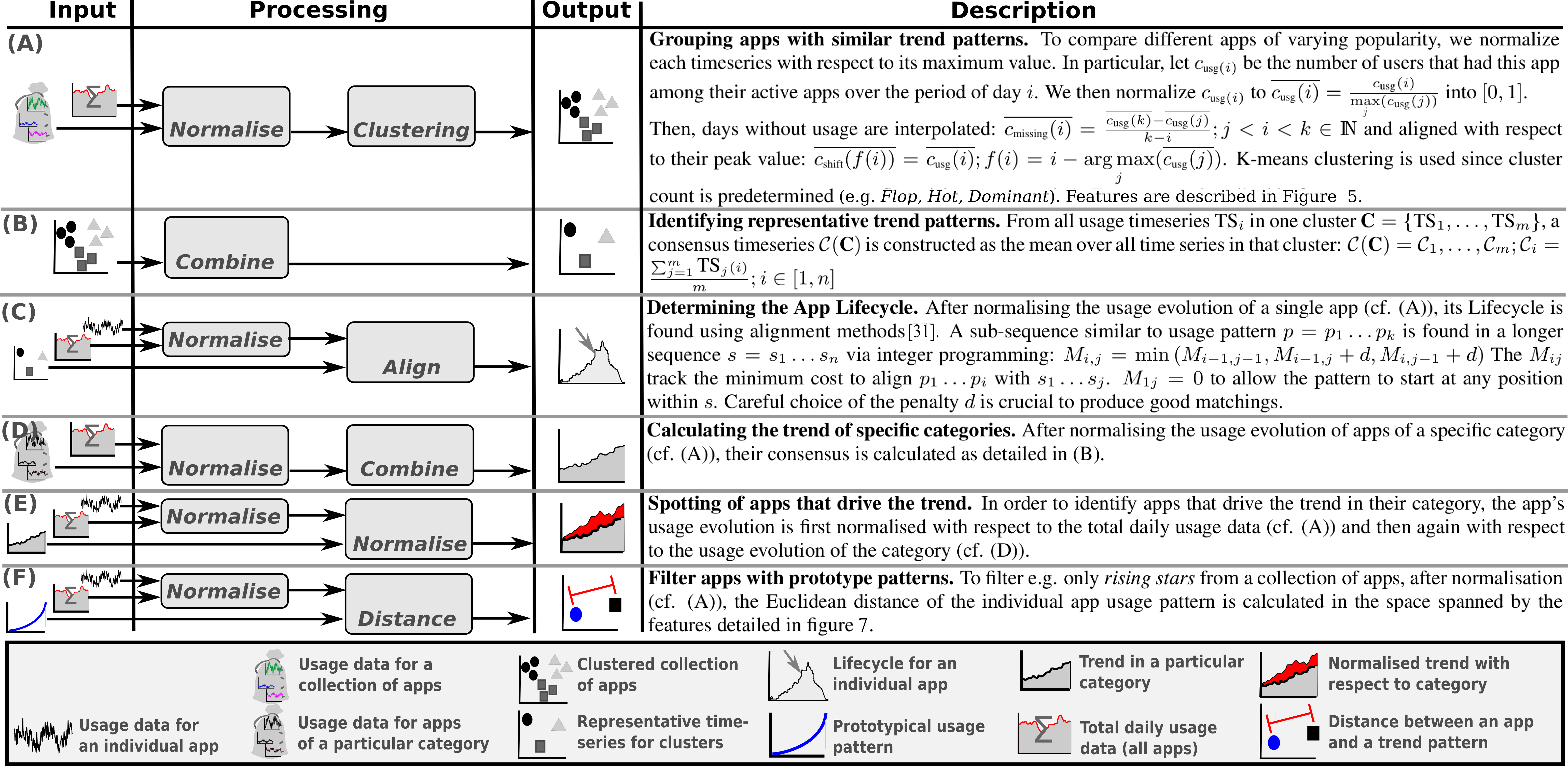}
 \caption{Description of a selection of trend mining procedures on real-time app usage data.}
 \label{figureOverview}
\end{figure*}
\nocite{4027}

We propose instantaneous application usage as a novel metric to accurately characterise the momentary hotness of mobile apps. 
Existing metrics, such as aggregated installation counts, co-installed apps and user reviews, often contain noise and biases making them an unreliable measure of an app's success or failure. 
Moreover, these measures ignore effects of temporal and other seasonal or external factors.
Figure~\ref{figureOverview} provides an overview of our toolbox for identifying and characterizing application usage trends. 
These cover (please refer to \fig{figureOverview} for details): 
\begin{description}
\item[(A) Group apps with similar trend pattern] Given prototype trend patterns (e.g. Dominant, Flop, Hot, Marginal (cf. \fig{figureLifecycles})), a set of apps is clustered according to their similarity to these prototypes. 
This module enables us to identify apps that resemble one of the four trend patterns.  
\item[(B) Extract representative patterns] From a group of apps, compute a consensus pattern as the mean over all patterns in the group. 
We use this module, for instance, to identifiy a representative pattern of a Google Play category by computing the consensus for all apps in the category (cf. (D)).
It can be applied to arbitrary groups of apps, for instance, to compare their average popularity (e.g. different groups of games or applications from a specific developer) 
\item[(C) Determining the App Lifecycle] An application can be at various stages inside their application lifecycle. 
For instance, beginning (initial stage), middle (rising), end (over the top). 
To find the stage of an app within a partiular lifecycle, we apply allignment approaches between the representative trend pattern and the app usage statistics. 
\item[(D) Calculate the trend of specific categories] To uncover the popularity trend of Google Play categories, we determine the consensus of the category to represent the overall trend.
\item[(E) Spot apps that drive the trend] We isolate dominant apps in a category that drive the usage trend. 
This is achieved by normalising the performance of the app against the performance of the category (using (B)). 
The resulting pattern displays the performance of the app with respect to all other apps in the same category. 
A positive slope in the pattern indicates that the app is performing better than its category while a negative slope indicates underperformance.
\item[(F) Filter apps with prototype patterns] We utilize the distance of an apps usage pattern to a prototype pattern in order to filter apps which closely resemble the prototype pattern. 
This module can be used, for instance, in analysis and recommendation of apps. 
In particular, since the usage trend is normalized against the trend of the category, the user or developer learns if the app is driving a trend or merely dominated by other apps in the category. 
\end{description}

\begin{figure}
  \includegraphics[width=\columnwidth]{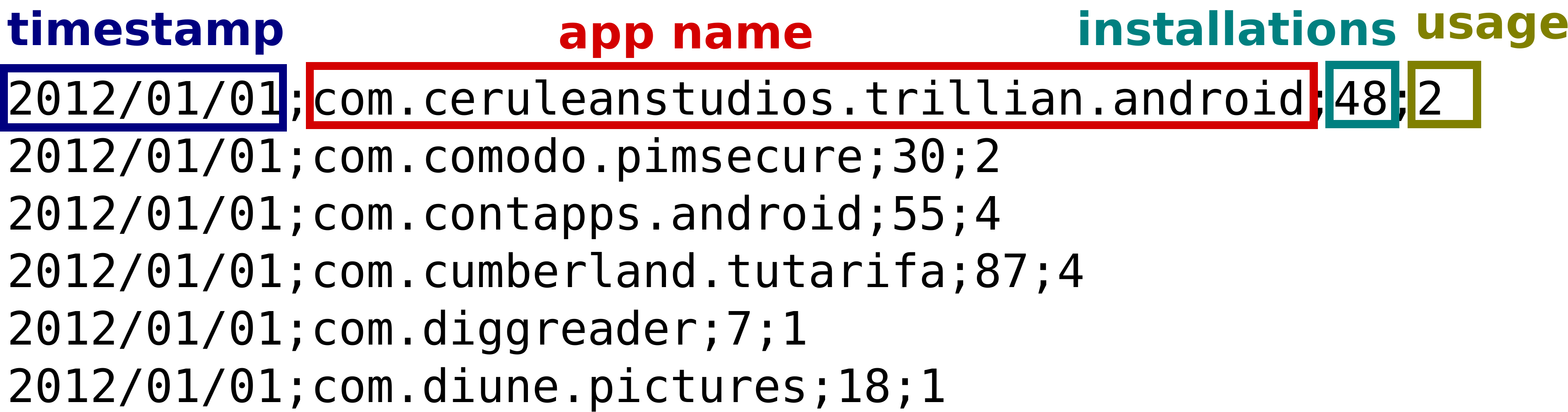}
 \caption{App-usage data from Android: process name, installation count, and daily usage.}
 \label{figureExampleData}
\end{figure}

For all above modules, the first step in the processing of real-time usage data is to extract from users interacting with a particular app on a specific day the usage frequency of individual apps (\fig{figureExampleData}), i.e., for each day, how often it was active across all mobile devices. 
Note that usage frequency is not proportional to installation count due to retention. 
%
%
%
%
The usage frequency data is potentially discontinuous (this is especially true for apps resenbling the {\em Marginal} pattern), and absolute values of different apps vary. 
To meaningfully compare app usage trends, normalization with respect to the total usage count (relative popularity over time) and within maximum usage of the app (popularity within a particular day) is therefore necessary.
In particular, for the number $c_{\mbox{\tiny usg}(i)}$ of users interacting with a particular app on day $i$, we normalize $c_{\mbox{\tiny usg}(i)}$ to $ \overline{c_{\mbox{\tiny usg}}(i)}=\frac{c_{\mbox{\tiny usg}}(i)}{\max\limits_j(c_{\mbox{\tiny usg}}(j))}$ into $[0,1]$.
Missing data is interpolated for discontinuous usage patterns.  
These normalised app usage patterns can then be exploited for comparison with other patterns and for further processing.

\subsection{Features and Clustering}
\begin{figure}
	\subfloat[Prototype trend patterns]{\includegraphics[width=.48\columnwidth]{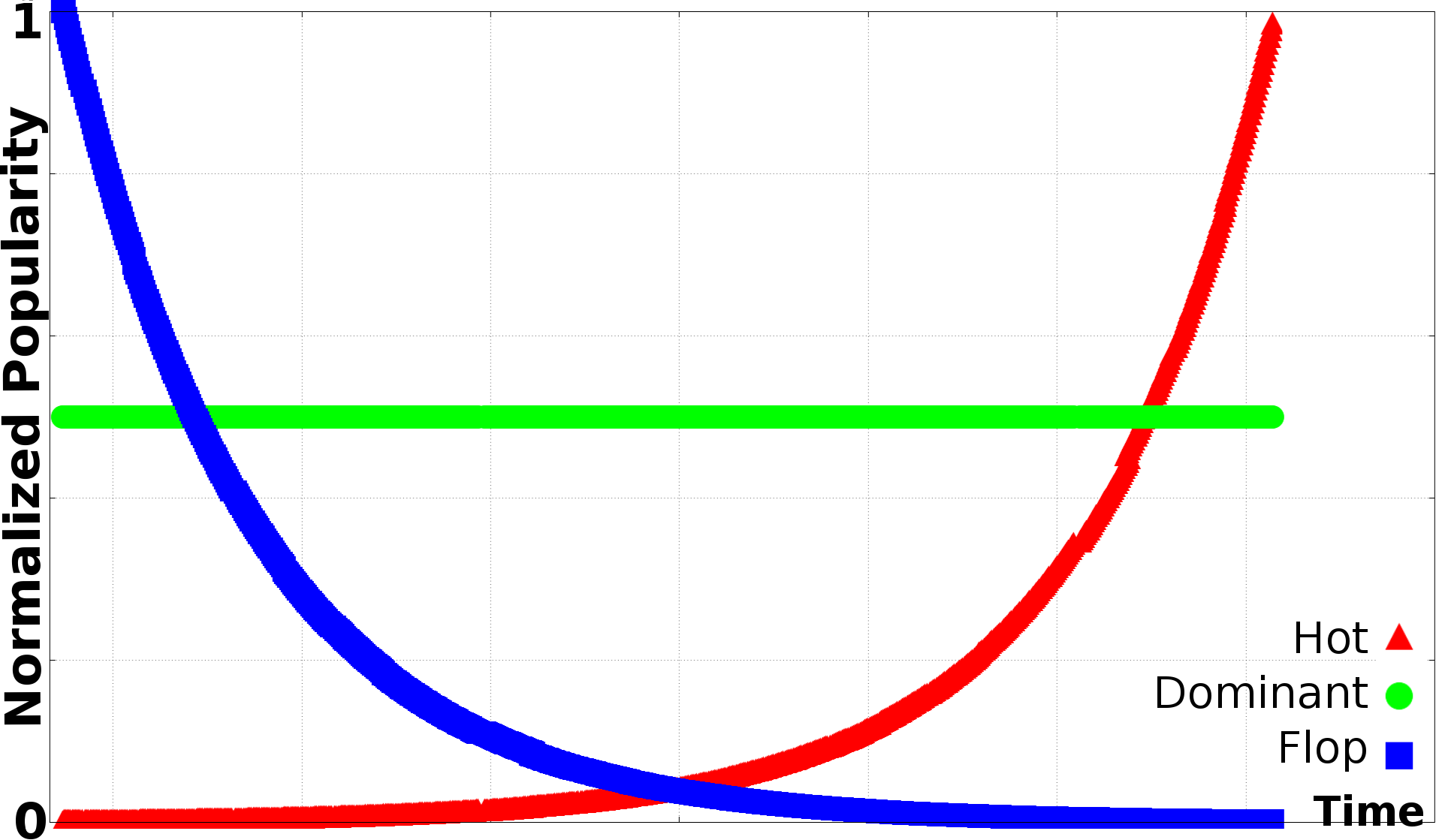} \label{figureLifecycleA}}\hfill
	\subfloat[Example Hot apps]{\includegraphics[width=.48\columnwidth]{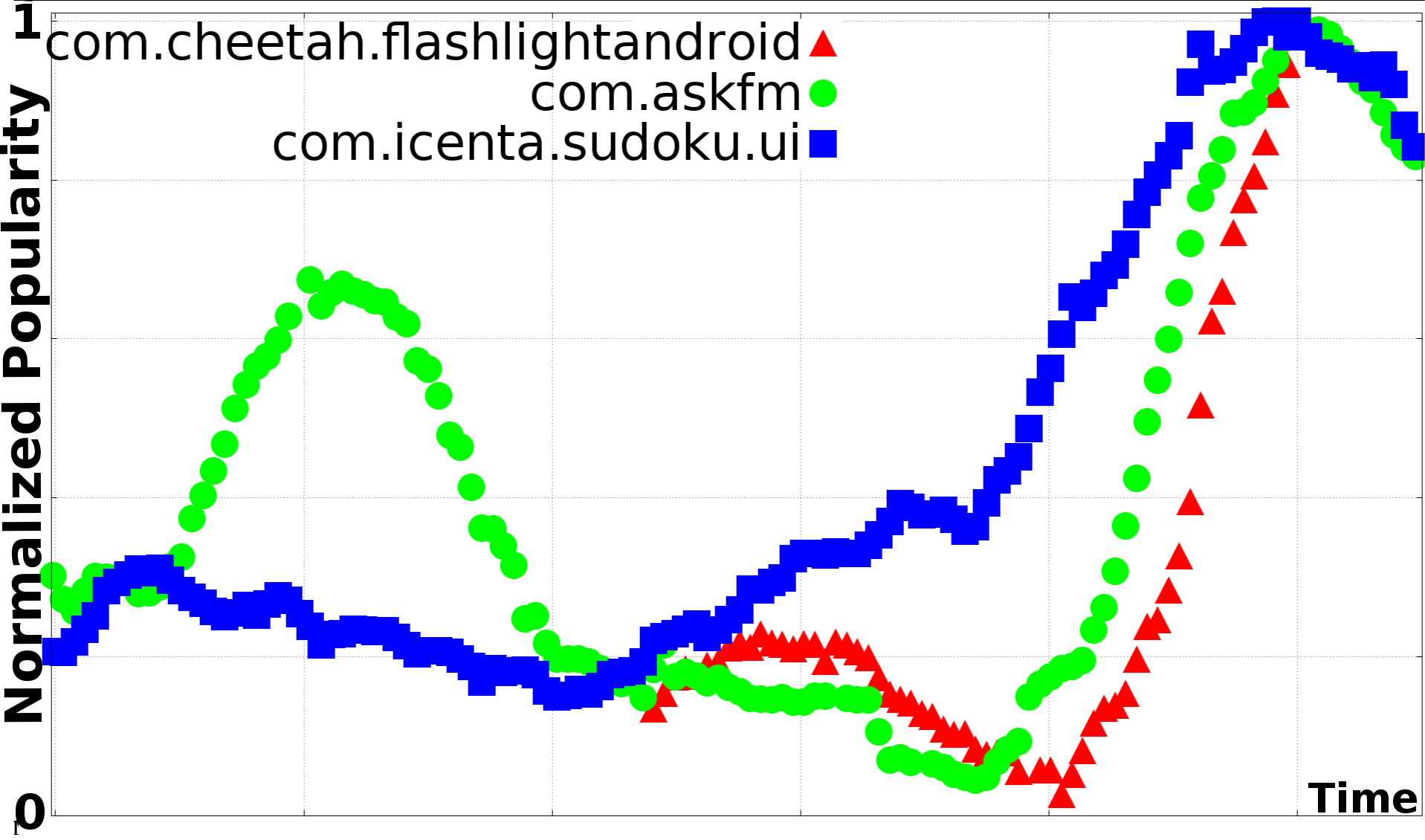} \label{figureLifecycleB}}
	
	\subfloat[Example Flops]{\includegraphics[width=.48\columnwidth]{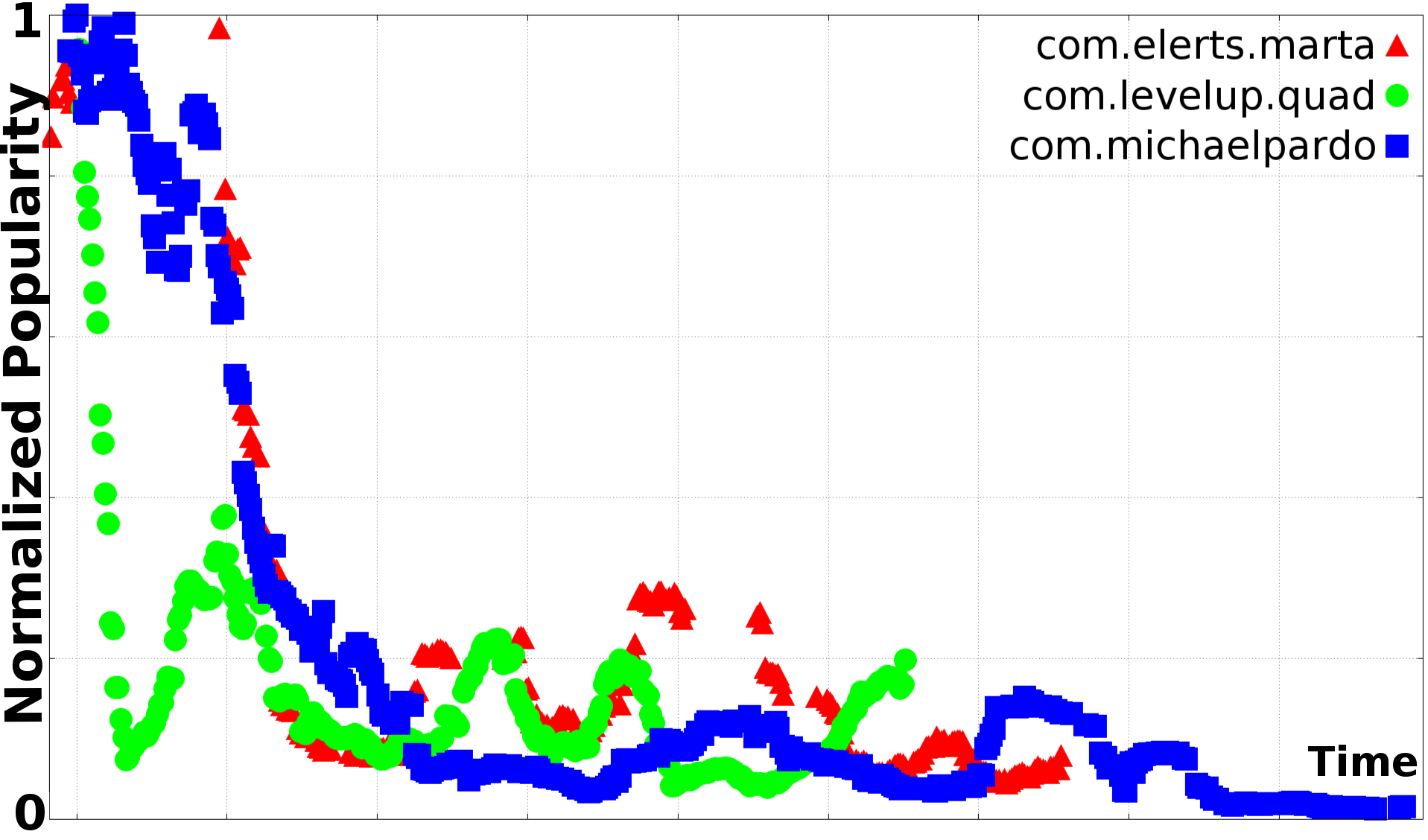} \label{figureLifecycleC}}\hfill
	\subfloat[Example Dominant apps]{\includegraphics[width=.48\columnwidth]{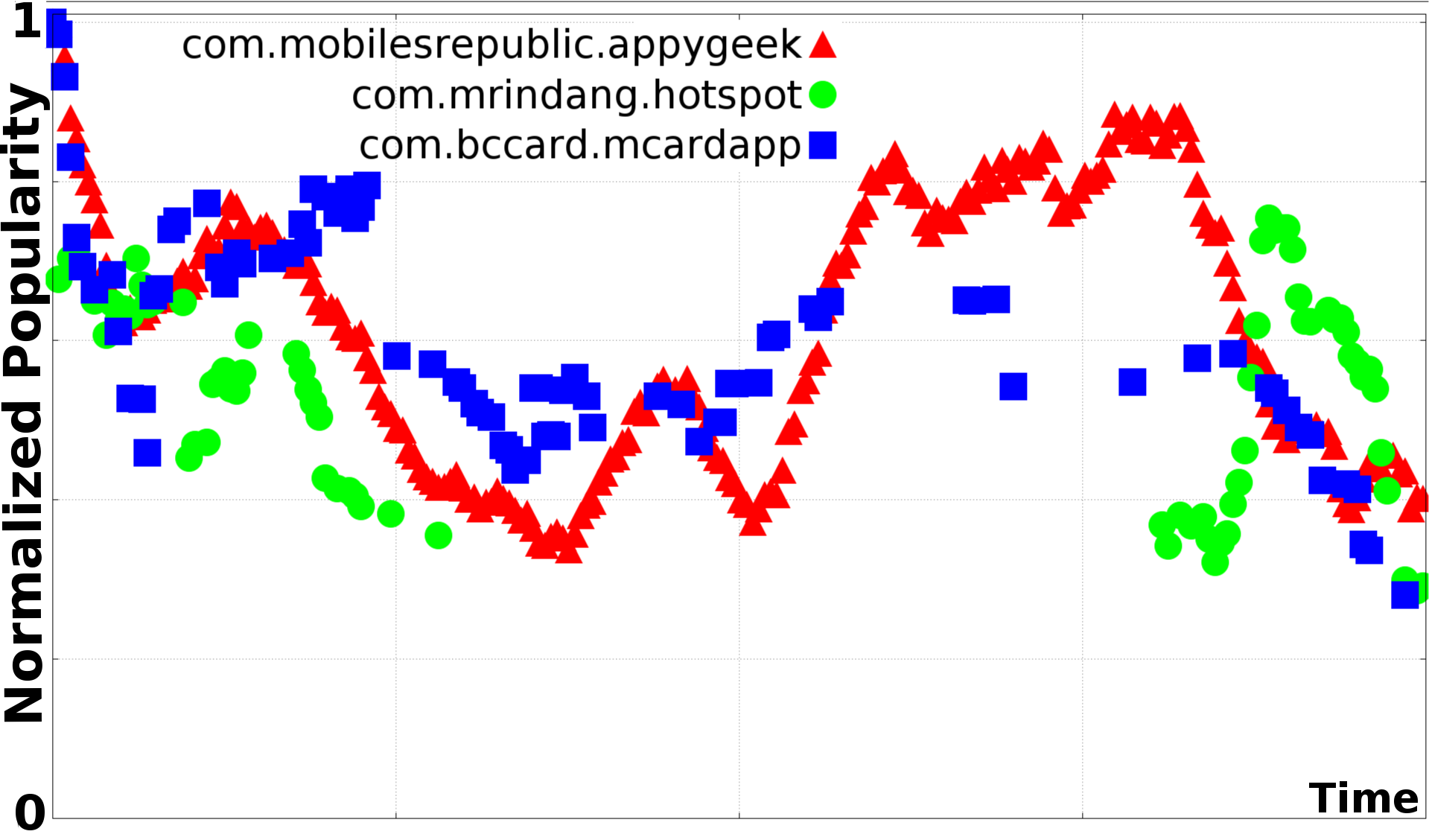} \label{figureLifecycleD}}
	\caption{App trend patterns with example apps.}
	\label{figureLifecycle}
\end{figure}
\begin{figure}
\subfloat[Distribution of the average size of clusters found]{\includegraphics[width=.48\columnwidth]{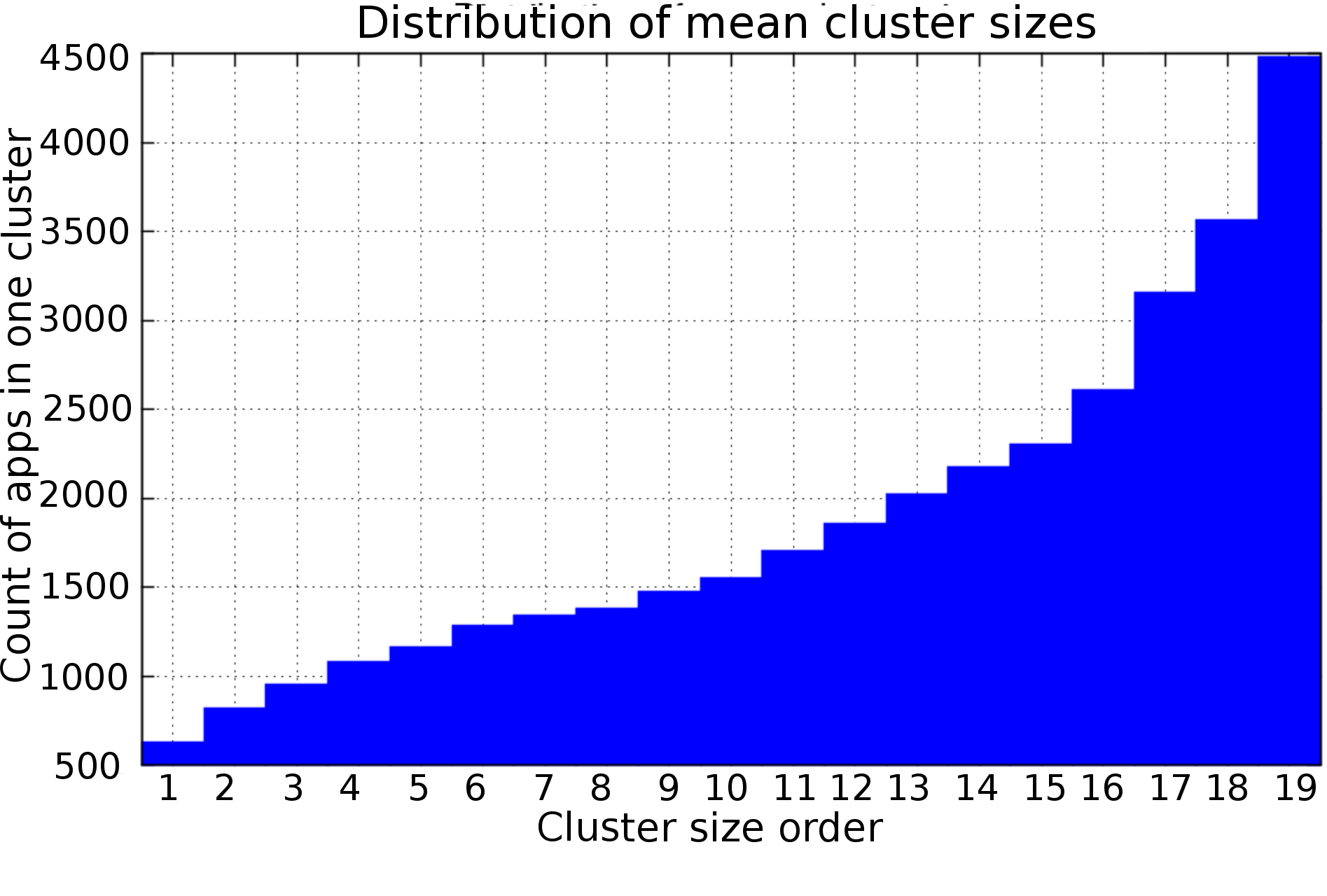} \label{figureTrendDistribution}}\hfill
\subfloat[Plot of the average evolution of the {\em Hot}, {\em Flop}, and {\em Dominant} patterns found in the clusters]{\includegraphics[width=.48\columnwidth]{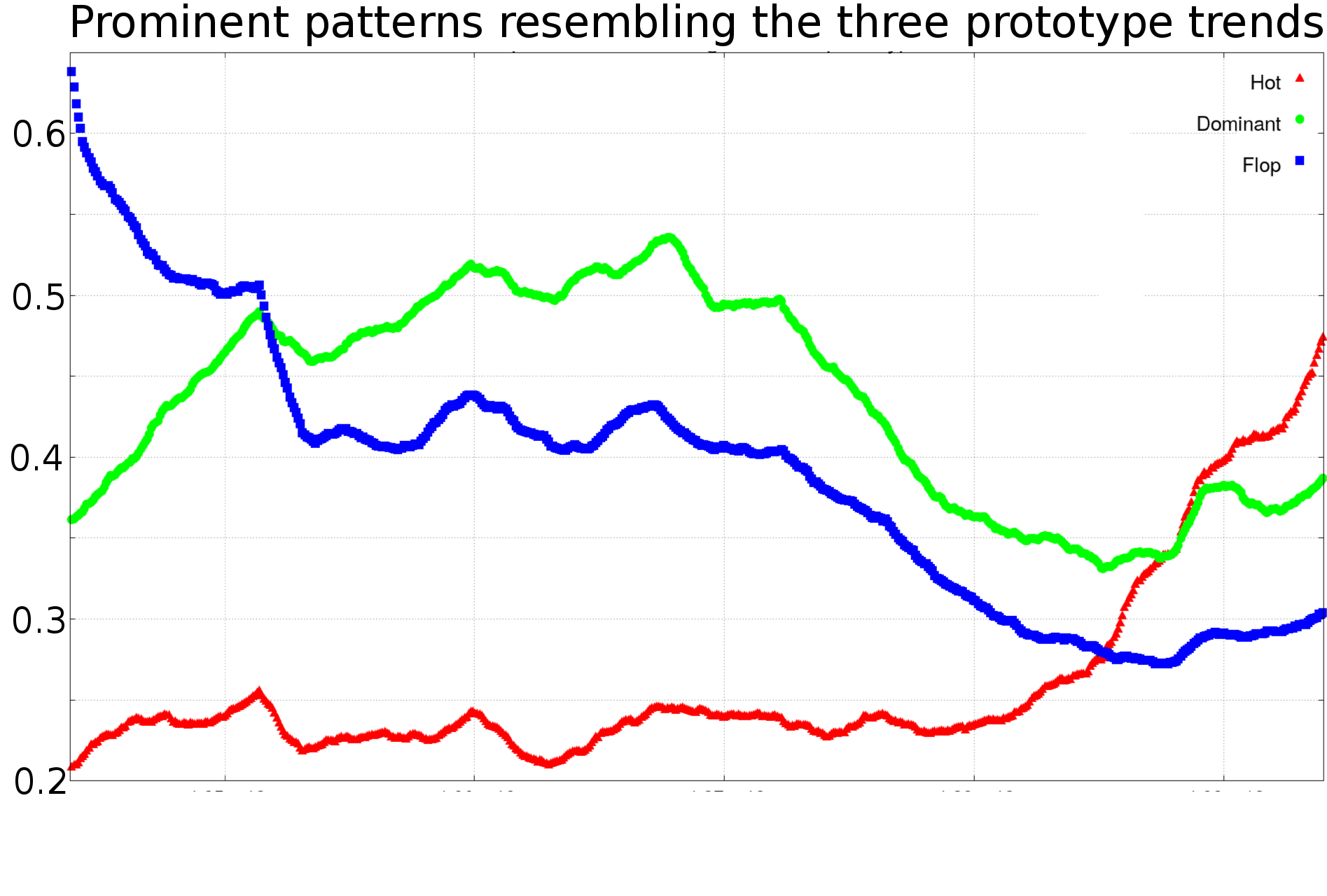} \label{figureProminentPatterns}}
 \caption{Distribution of cluster sizes and exemplary prototype clusters found among the clusters}
\end{figure}

Individual apps follow distinct instantaneous trends. 
This is depticed in \fig{figureLifecycle} for a selection of exemplary apps after normalization. 
We are interested in identifying the four main trends 'rising in popularity' ({\em Hot}), 'falling in popularity' ({\em Flop}), 'constant high popularity' ({\em Dominant}) and 'no popularity' ({\em Marginal}) (cf. figure~\ref{figureLifecycleA}).
Although there are more and fine-grained trends to be found in the data, we focus in our study on these four, since they are found prominently in the data and have the potential to imply clear recommendation for both app developer and app user. 
In particular, figure~\ref{figureTrendDistribution} depicts the distribution of mean cluster sizes we found after 10 runs of k-means with $k=20$.
The {\em Marginal} pattern was constanty found by about one order of magintude more often than all other patterns (not shown in the figure). 
The three main trends {\em Hot}, {\em Flop}, and {\em Dominant} have been prominently found among the clusters computed in all runs of the k-means algorithm (cf. for instance, figure~\ref{figureProminentPatterns}). 
As visible from figure~\ref{figureTrendDistribution}, on average, the three largest clusters jointly represent about $\frac{1}{3}$ of the apps with a long tail of small clusters. 
This shows that, although more than the {\em Hot}, {\em Flop}, and {\em Dominant} (and {\em Marginal}) patterns can be found in the data, their relevance is rather small. 
We leave the study of further clusters to future work and focus in this paper on the prototype clusters identified in figure~\ref{figureLifecycleA}.

\fig{figureLifecycleB}, \ref{figureLifecycleC} and \ref{figureLifecycleD} depict exemplary app usage evolutions for specific apps filtered with one of these prototype trend patterns (cf. Module (F) in \fig{figureOverview}).
While more complex patterns can also be found with the proposed modules, these are combinations of the above patterns. 
Furthermore, we are mostly interested in the instantaneous trend of the application and less in historical data.
We expect that the former three basic trends will approximately resemble the three characteristic patterns illustrated in \fig{figureLifecycleA}. 
Apps with very low usage ({\em Marginal} pattern) closely resemble $f(x)=1$, a straight line with constant value '1'. 
This is due to our pre-processing of applications in order to make them comparable:  
Normalisation with respect to the highest observed daily use followed by the interpolation of missing values. 

\begin{figure}
\includegraphics[width=\columnwidth]{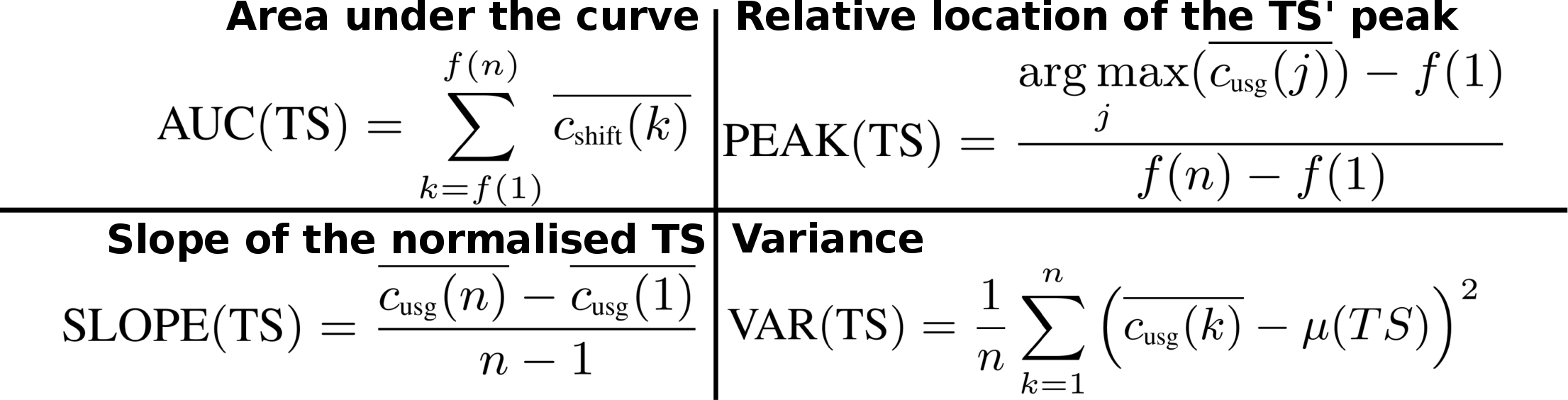}
\caption{Features exploited for an app usage time series $\mbox{TS}=\overline{c_{\mbox{\tiny shift}}(f(1))},\overline{c_{\mbox{\tiny shift}}(f(2))},\dots,\overline{c_{\mbox{\tiny shift}}(f(n))}$ to judge the similarity of their shape. $\mu(TS)$ is the arithmetic mean of the timeseries. }
 \label{figureFeatures}
\end{figure}

In order to compare normalized usage patterns of applications to each other, to consensus patterns of groups of apps or to trend patterns as specified in \fig{figureOverview}, we measure similarity via Euclidean distance in a feature space spanned by {\em Area Under the Curve} (AUC), {\em Relative Peak location} (Peak), {\em Slope} and {\em Variance} (cf. \fig{figureFeatures}). 
The choice of these features is owed to the nature of the four trend patterns to distinguish. 
{\em Peak} and {\em Slope} are able to distinguish {\em Hot} or {\em Flop} patterns, whereas the {\em AUC} distinguishes those from the other two constant patterns which have a much larger {\em AUC}.
Finally, {\em Variance} and {\em AUC} are able to distinguish between {\em Dominant} and {\em Marginal} as the latter will have low {\em variance} and high {\em AUC}.

With these features, trend patterns are clustered with k-means clustering according to their similarity to a the four main trend patterns (\fig{figureOverview} (A).
We utilise k-means since the number of clusterheads is fixed by the four trend patterns {\em Hot, Flop, Dominant}, and {\em Marginal}. 

\subsection{Impact of App Categories}
\begin{figure}
\subfloat[Trend evolution of 3 categories from Google Play Store]{\includegraphics[width=.52\columnwidth]{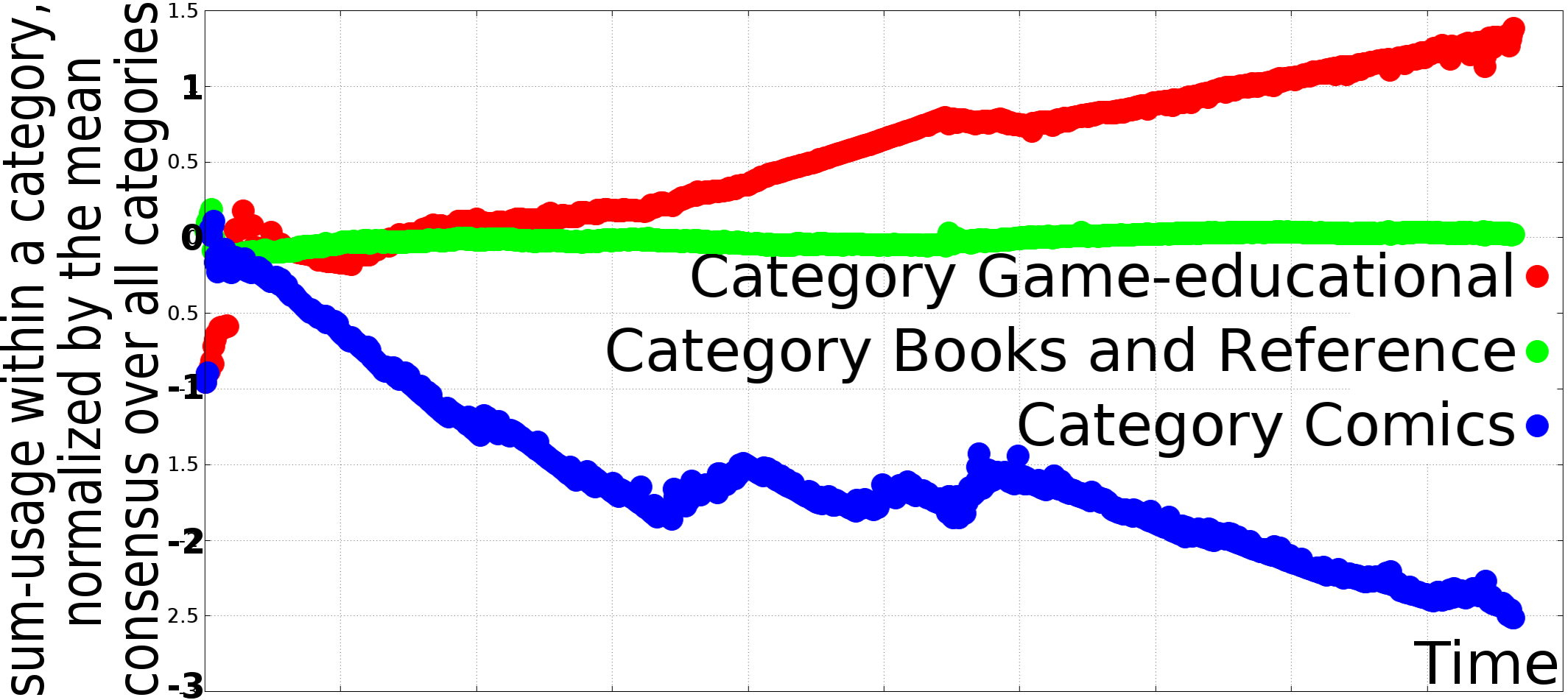}\label{figureCategoriesComparison}}
\hfill \subfloat[Facebook dominates the category Social]{\includegraphics[width=.47\columnwidth]{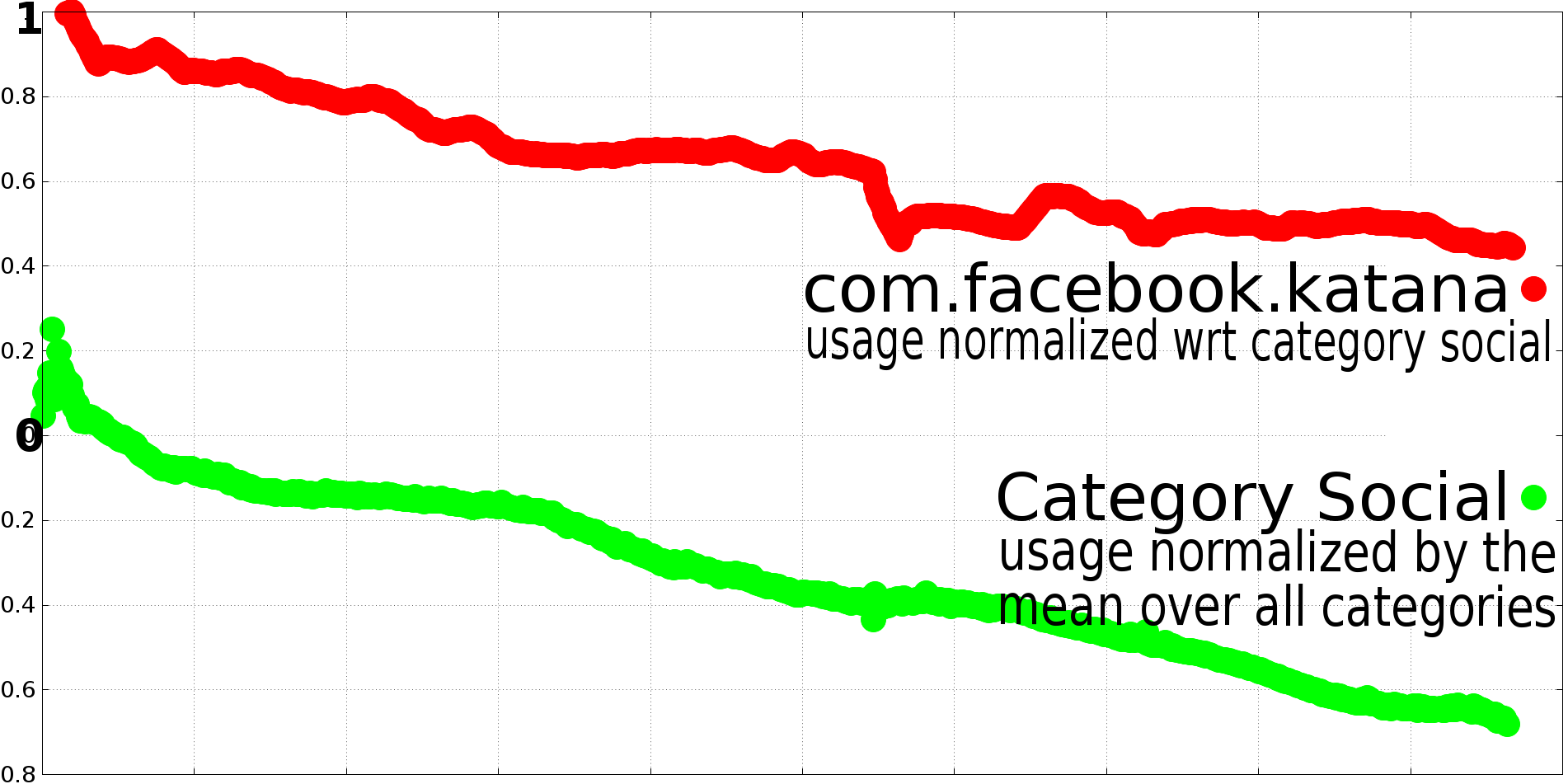}\label{figureCategoriesDominating}}
\caption{Trend evolution of categories and individual apps}
 \label{figureCategories2}
\end{figure}

In trend analysis, we are interested in the popularity of specific apps with respect to others. 
However, apps can fall into various categories and might not be comparable, such as, for instance, categories {\em Game} and {\em Business}. 
In particular, in the Google Play Store, some categories might increase in popularity while others decrease. 
This can be seen in \fig{figureCategoriesComparison} where the trend of the exemplary categories {\em Game-educational}, {\em Books and Reference}, and {\em Comics} is plotted (cf. (D) 'Calculating the trend of specific categories' in \fig{figureOverview}). 
The displayed consensus trend of the categories shows distinct slope. 

Overly successful apps have a significant impact on the trend of their category. 
An example is the impact of Facebook on its category {\em Social}. 
The Facebook app dominates the performance of the category as indicated in \fig{figureCategoriesDominating}. 
Compared to the consensus of all apps in this category, {\em com.facebook.katana} features with 0.2105 the smallest Euclidean Distance in the feature space. 
Smaller copycat apps branch part of the success and popularity of dominant ones. 
Due to the success of popular apps like, for instance, {\em WhatsApp}, many other similar apps are published and users attracted to this category not only follow the leading app but also try other, similar, possibly localized, alternatives, such as WeChat (Weixin) in China.  
Smaller apps might then appear to follow a rising trend but merely benefit from the overall popularity of the category. 

For a fair comparison of apps that belong to different categories, the overall trend of the categories should therefore be subtracted out.

\begin{figure*}
\subfloat[Consensus of the cluster, describing the average popularity of apps in this cluster]{\includegraphics[width=.32\textwidth]{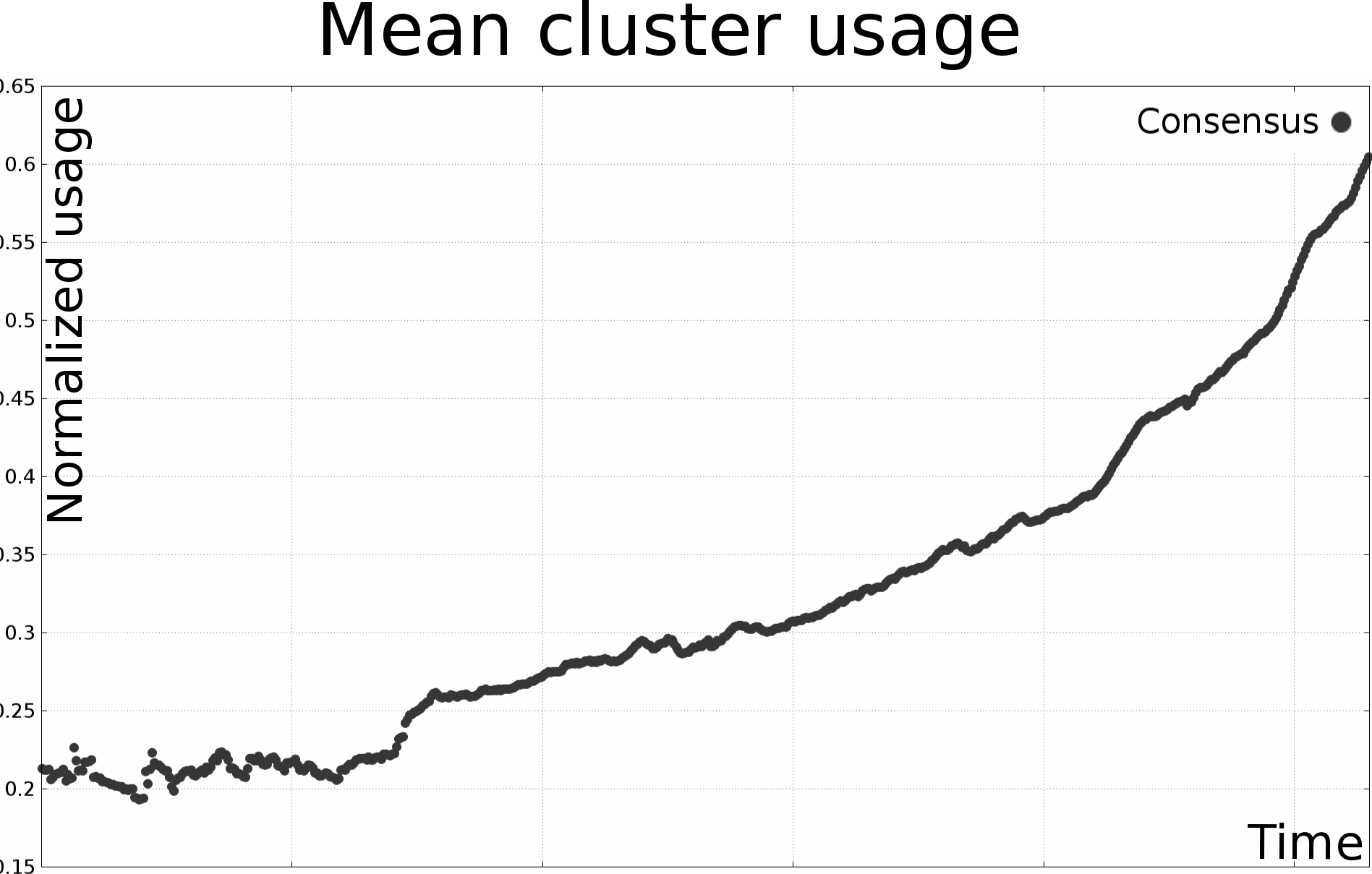} \label{figureRelativePerformanceA}}\hfill
\subfloat[Normalised performance of three individual apps from this cluster]{\includegraphics[width=.32\textwidth]{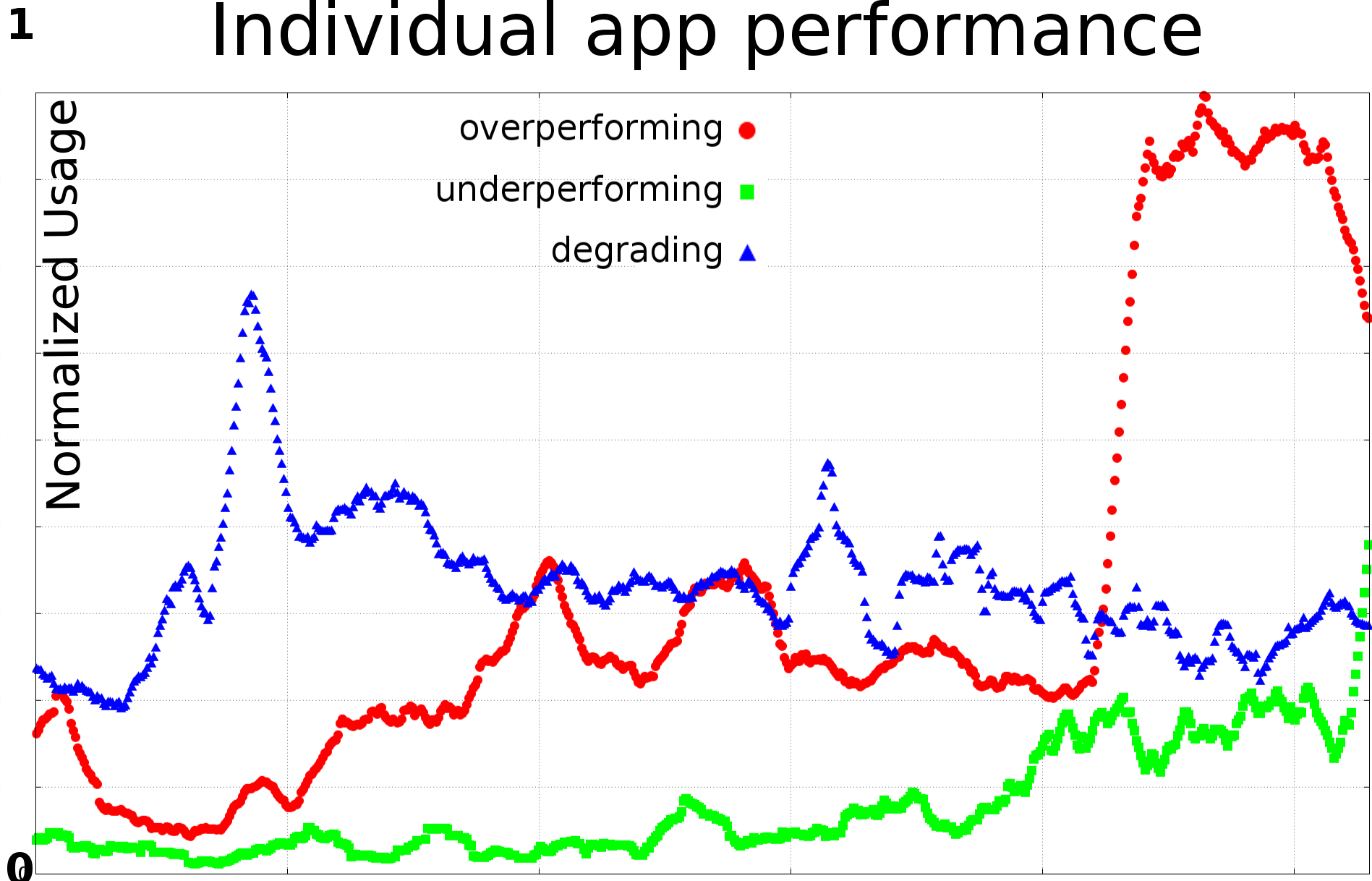} \label{figureRelativePerformanceB}}\hfill
\subfloat[Relative performance of the apps with respect to the cluster]{\includegraphics[width=.34\textwidth]{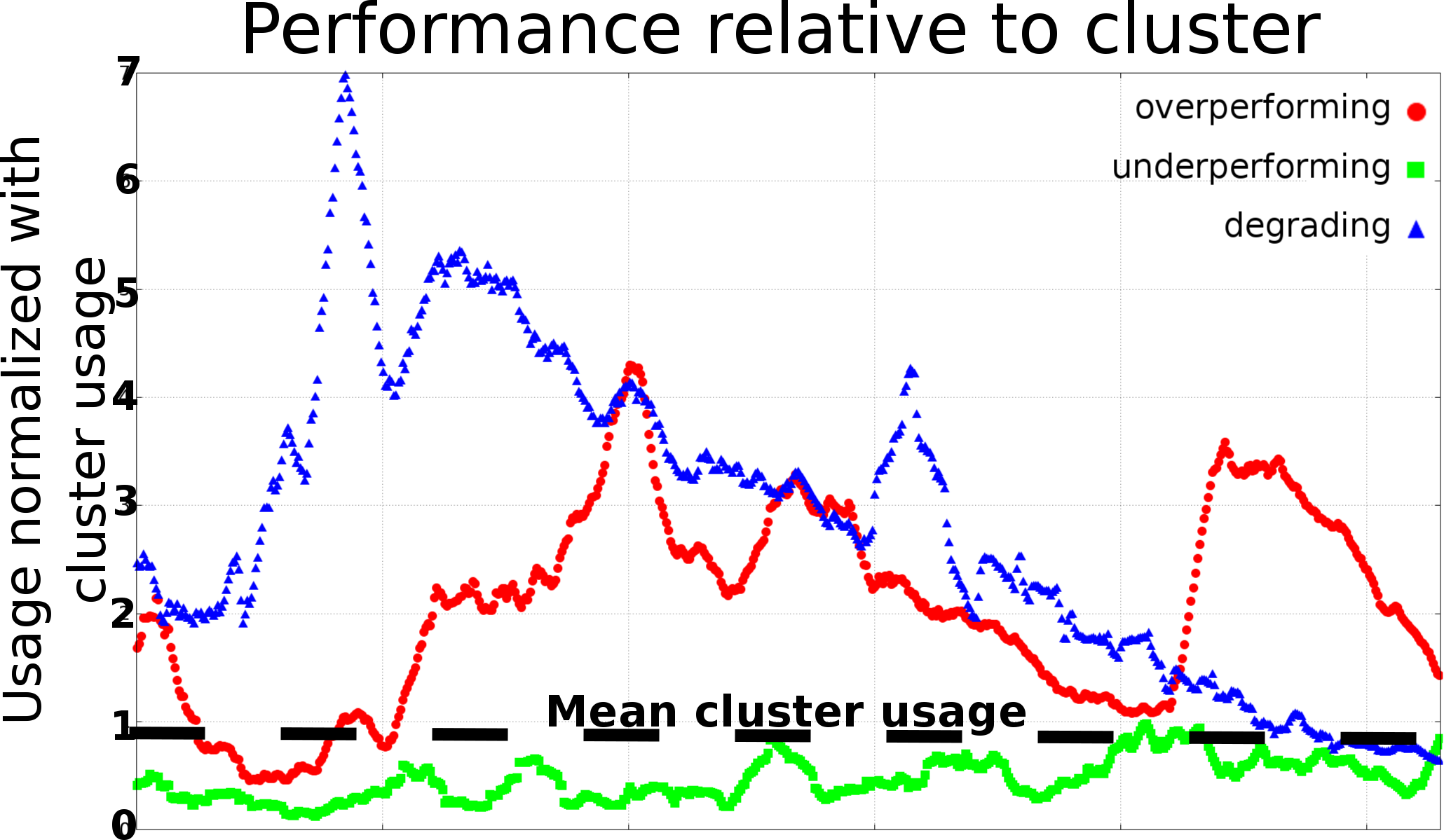} \label{figureRelativePerformanceC}}
\caption{Performance of individual apps relative to the performance in their cluster.}
\label{figureRelativePerformance}
\end{figure*}

For recommendation or analysis purpose, apps that drive the trend are especially interesting over copycat-apps which are worse than the trend. 
To objectively measure the performance of an app, free of the influence of its category, we calculate its performance relative to the performance of its category (cf. {\em (E) Spotting apps that drive the trend} in \fig{figureOverview}).
To illustrate that normalisation against the overall trend in a category is beneficial to identify the actual instantaneous performance of an app consider \fig{figureRelativePerformance} for three exemplary apps in one category. 
\fig{figureRelativePerformanceA} plots the consensus of the app category.
Looking at individual app's performance (\fig{figureRelativePerformanceB}), their trend is hardly visible.
However, after normalization with the cluster's consensus pattern in \fig{figureRelativePerformanceC}, the app represented with \color{red}{\huge $\bullet$}\color{black} is overperforming as it regularly scores above the cluster's performance while the app labelled \color{green}$\blacksquare$\color{black} is constantly underperforming. 
Finally, the third app (labelled \color{blue}$\blacktriangle$\color{black}) is dominating in the beginning while then constantly degrading in performance. 
Observe that, in contrast to this normalized evolution, considering the apps individually, as in \fig{figureRelativePerformanceB}, the underperforming app actually appears to be rising in popularity while the degrading app appears to remain stable. 

\section{Empirical Evaluation}
\label{sec:results}
To demonstrate the value of the trend filter, we carried out experiments using the Carat dataset. 
We first show that our trend filter is capable to identify known success and failure storie.
Afterwards, we present a categorization of trends within different application categories on Google Play. 

\subsection{Validation}
\begin{figure}
\includegraphics[width=\columnwidth]{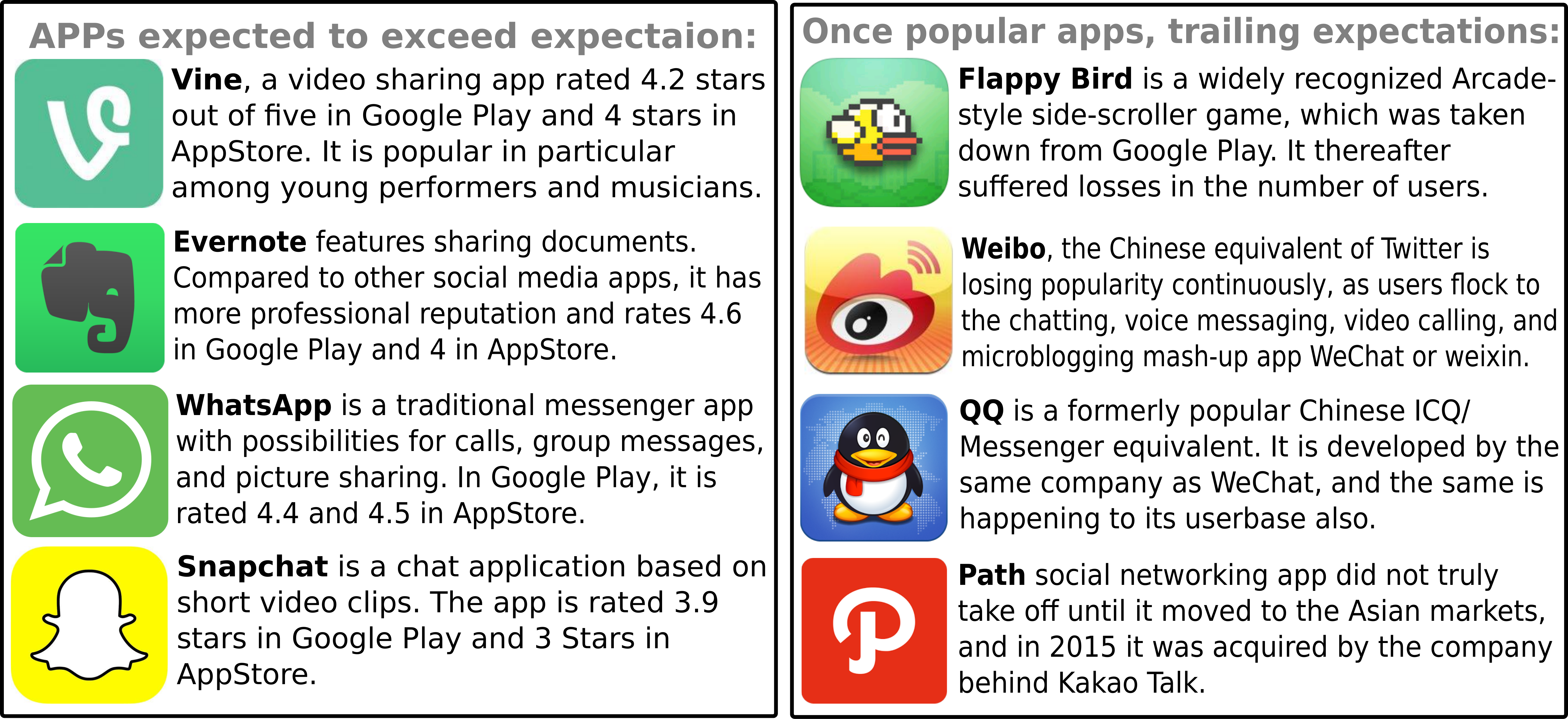}
\caption{Exemplary popular apps chosen for comparison}
\label{figurePopularApps}
\end{figure}

We begin our evaluation by demonstrating that trends can detect known phenomena, such as widely known popular and unpopular apps. 
We examine eight popular apps (cf. figure~\ref{figurePopularApps}), four of which are widely believed to be exceeding expectations and four which have been extremely popular once but which have significantly dwindled since. 

After grouping these apps according to the four representative trend patterns (cf. \fig{figureOverview}, module (A)), three of the popular apps, Vine, Evernote, and Snapchat are classified as {\em Hot} apps, whereas Whatsapp is classified as a {\em Dominant} type. 
Table~\ref{tab:success-flop-summary} illustrates the distance of each app to the nearest of the four representative trend patterns. 
From the table we observe that the distance from the {\em Hot} trend pattern for SnapChat is 25\% higher than for the other apps. 
This reflects also the lower app-store ratings it has received compared to the other apps. 
\begin{table}
	\caption{Categorization of example applications.}
	\label{tab:success-flop-summary}
	\centering
	\begin{tabular}{l|c|c|c}
		App & Category & Pattern & Distance\\\hline
		Vine & Entertainment & {\em Hot} & 0.4220\\ 
		Evernote & Productivity & {\em Hot} & 0.3956\\ 
		Snapchat & Social & {\em Hot} & 0.5399\\ 
		Path & Social & {\em Dominant} & 0.4467\\ 
		WhatsApp & Communication & {\em Dominant}& 0.1186\\ 
		Flappy Bird & Game - Arcade & {\em Flop} & 0.0575\\ 
		Weibo & Social & {\em Flop} & 0.2343\\ 
		QQ & Social & {\em Flop} & 0.2854\\ 
	\end{tabular}
\end{table}

\begin{figure*}
	\subfloat[Trend evolution of Snapchat]{\includegraphics[width=0.245\textwidth]{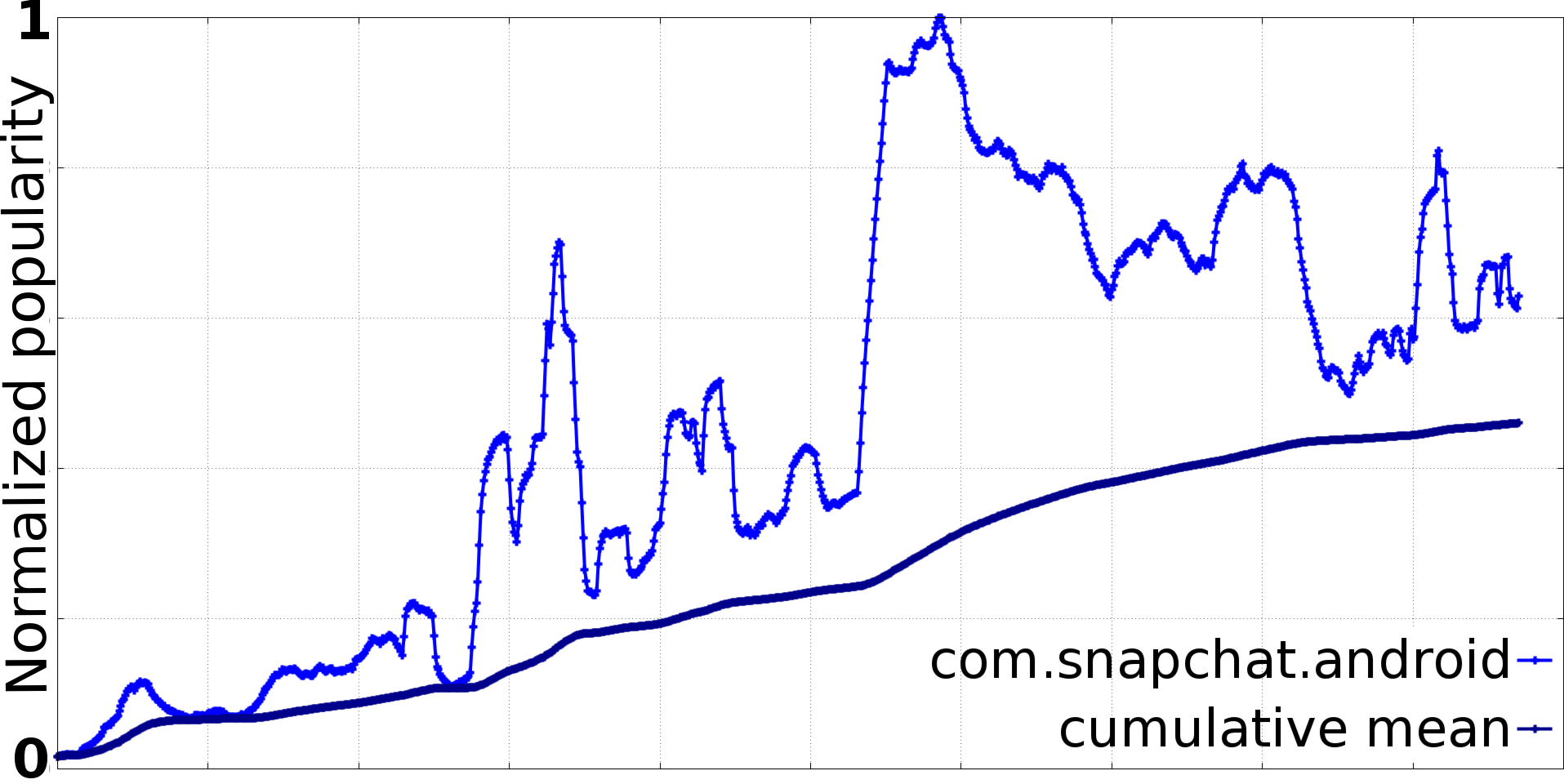}
	\label{figureSuccessStoriesA}}\hfill
	\subfloat[Trend evolution of Evernote]{\includegraphics[width=0.245\textwidth]{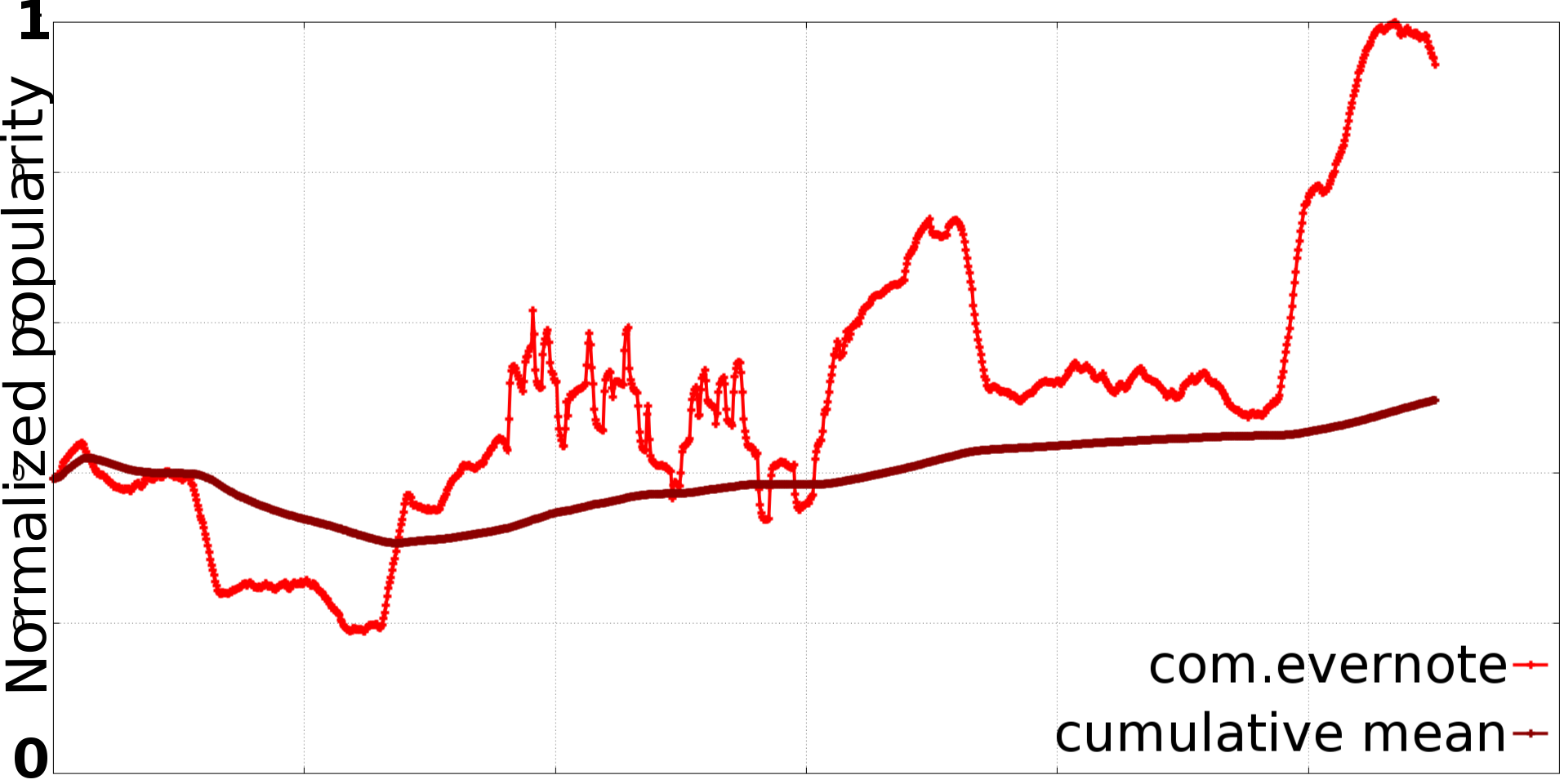}
	\label{figureSuccessStoriesB}}\hfill
	\subfloat[Trends of Flappybird and Path]{\includegraphics[width=0.245\textwidth]{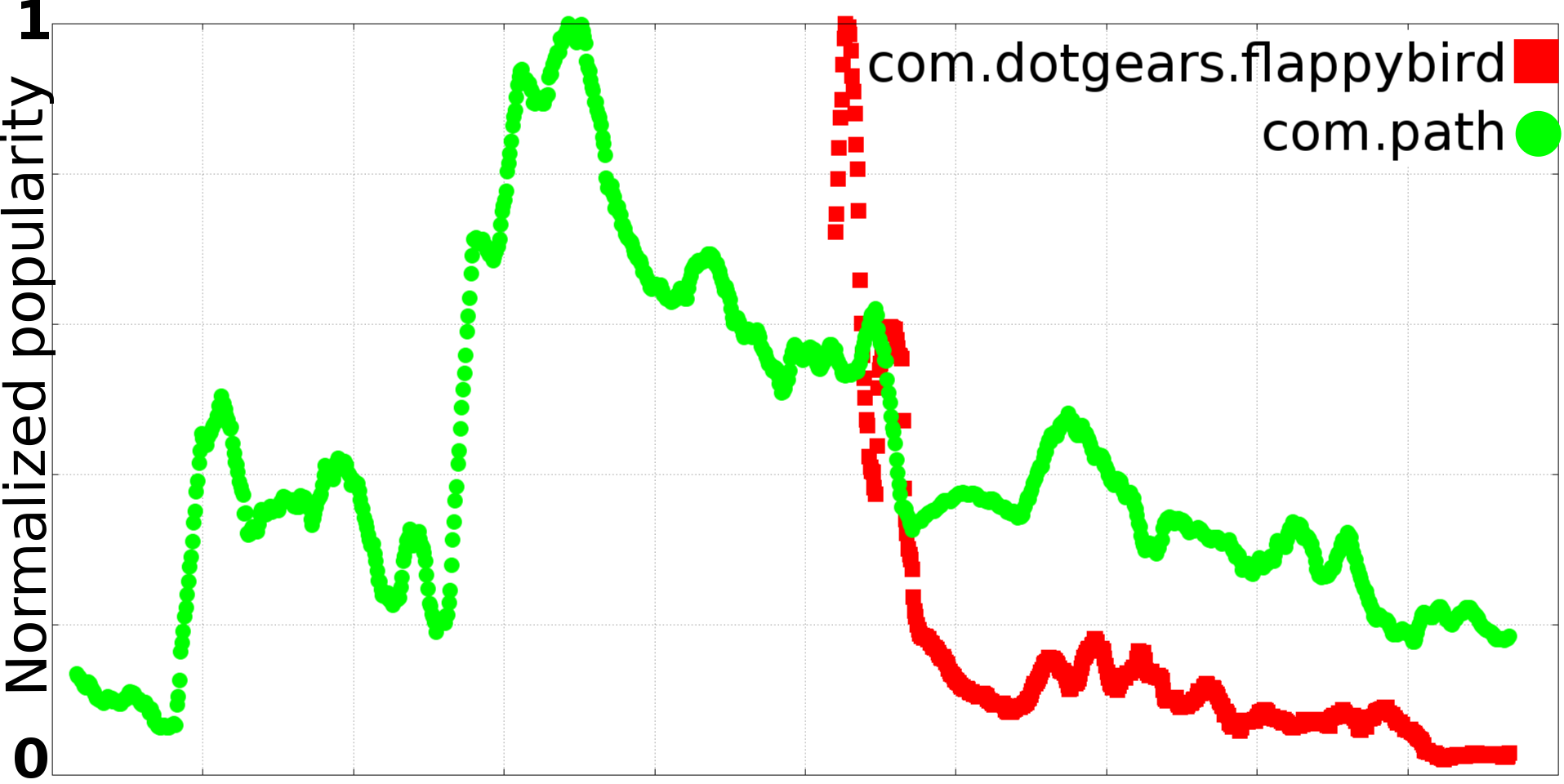}
	\label{figureFloppingApps}}\hfill
	\subfloat[Angry Birds trend-correlation]{\includegraphics[width=.245\textwidth]{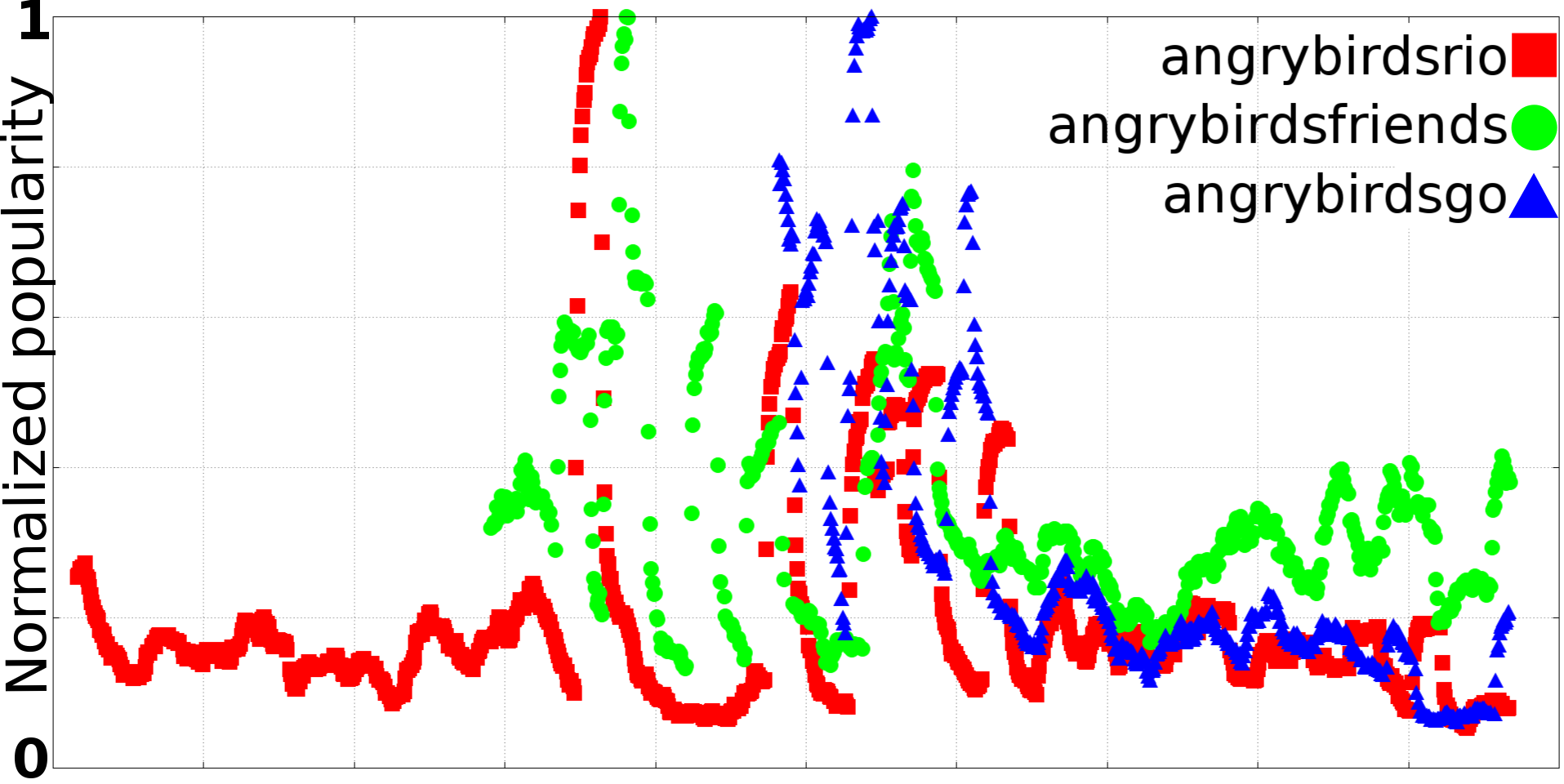}
	\label{figureAngrybirdsA}}
	\caption{Trends and common patterns in example app trend patterns}
\end{figure*}

We can see this also in the trend pattern the app experiences as compared to the other apps (cf. \fig{figureSuccessStoriesA} and \fig{figureSuccessStoriesB}).  
We can see that SnapChat features a steadily rising cumulative mean but also exhibits up and down fluctuations which affect its categorization. 
Evernote, in comparison, achieves a more stable long-term popularity and features a steep raise in its more recent trend. 
Nevertheless, our analysis is able to correctly identify a positive trend for each of these apps.

From the underperforming apps only three ({\em Flappy Bird}, {\em Weibo} and {\em QQ}) are classified as Flops. 
From the distances in Table~\ref{tab:success-flop-summary}, we observe that {\em Flappy Bird}, {\em Weibo} and {\em QQ} are very close to the archetypical flop pattern, whereas the distance of Path to the Dominant pattern is high, suggesting that it is still in transition between lifecycle states.
The graphical comparison between {\em Flappy Bird} and {\em Path} in \fig{figureFloppingApps} stresses this observation: 
While {\em Flappy Bird} follows closely an architypical Flop-pattern, the evolution of {\em Path} has returned to past levels after a temporary high.

It is important to stress here that the Flop pattern should not be interpreted as negative. 
On the contrary, it indicates that the app was successful to gather a huge user base in short time, but experienced high loss in users quickly thereafter, i.e., it has low retention rate. 
As discussed in Section~\ref{sectionRetentionRate}, it is a natural matter of retention rate that the usage drain is significant, and Flop simply means that the app has surpassed its 'best before' date. 
As an example, consider the Angry Birds series of apps. 
As indicated in Figure~\ref{figureAngrybirdsA}, the patterns of most Angry Birds apps closely resemble the Flop pattern (shifted by their respective release date) even if most of them can be considered to be exceptionally successful.  
Furthermore, we can observe that a series of apps in the same theme or product family has the potential to benefit each other as a compounding effect. 
We see in Figure~\ref{figureAngrybirdsA} how the much older {\em Angry Birds Rio} also experiences a rise in popularity at the time the {\em Angry Birds Friends} and {\em Angry Birds Go} are released. 
These peaks can also be observed at very similar times for the original {\em Angry Birds} and the older {\em Angry Birds Seasons}.

\subsection{Large-Scale Study of Application Trends}
We next analyse the distribution of {\em Flop}, {\em Hot} and {\em Dominant} apps (\fig{figureLifecycleA}) within various categories. 
Throughout all categories, about 40\% or more of the apps are marginal. 
This means that from all apps in Google Play, less than 60\% ever gather more than a handful of users.
Table~\ref{tableFlopTopHigh} summarizes the frequency of the four main trend patterns in exemplary Google Play categories.

\begin{figure*}
\includegraphics[width=\textwidth]{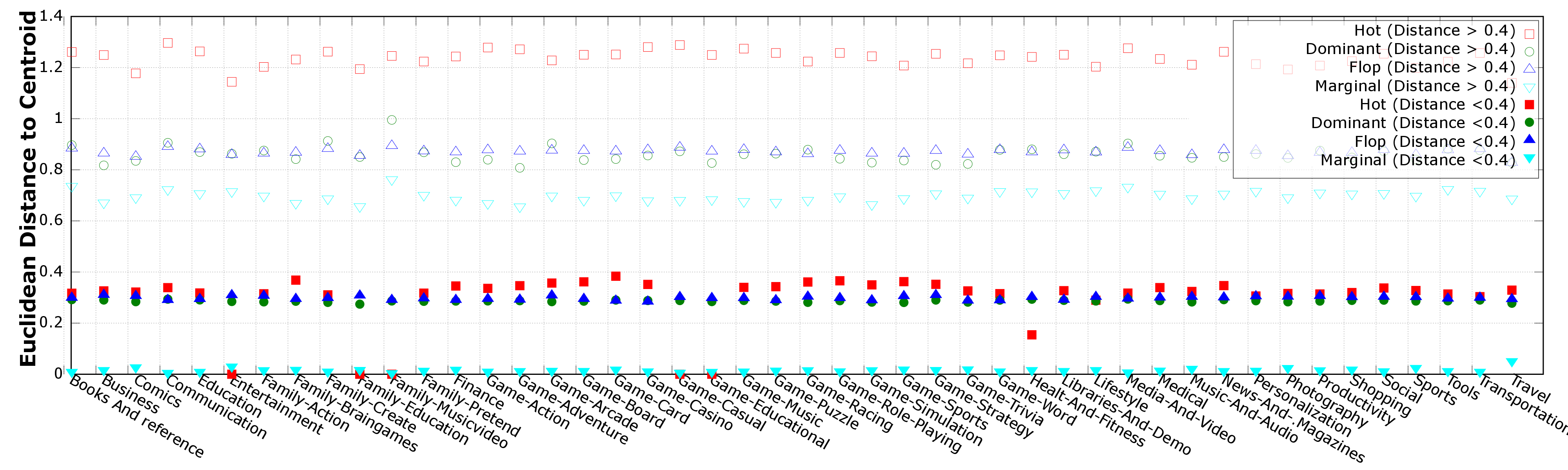}
	\caption{Mean Euclidean distance to the respective cluster centroid ({\em Hot}, {\em Dominant}, {\em Flop}, {\em Marginal}) for apps associated with the respective trend pattern (low Euclidean Distance) compared to those not associated with it (high Euclidean Distance)}
		\label{figureClusterStatistics}
\end{figure*}

We cluster the remaining 60\% to a particular trend pattern when the Euclidean Distance to the respective cluster centroid ({\em Hot}, {\em Flop}, {\em Marginal} or {\em Dominant} patterns) is below $0.4$.
In our tests, this distance was able to clearly separate the apps belonging to one of the four patterns from those that do not belong to it (cf. \fig{figureClusterStatistics}). 
In the figure, we have calculated the mean Euclidean distance of groups of apps to their nearest cluster centroid. 
The figure compares the mean distance of apps that are associated with a respective trend pattern to those that feature an Euclidean distance greater than 0.4. 

Observe from table~\ref{tableFlopTopHigh} that for the apps associated with one of the trend patterns, the {\em Marginal} apps are most similar to their respective trend pattern and the mean Euclidean distance for {\em Hot}, {\em Dominant} and {\em Flop} apps is for all categories sharply concentrated around 0.3. 
The distance of the remaining apps to their nearest trend pattern is significantly higher. 
This means that they might follow other, more complex patterns or experience constant fluctuation.
We leave the study of further, more complex trend patterns to also represent these apps open for future research.

\begin{table*}
\caption{Percentage of {\em Marginal}, {\em Flop}, {\em Dominant}, and {\em Hot} apps for 17 exemplary categories}
\label{tableFlopTopHigh}
\centering
\begin{scriptsize}
\begin{tabular}{r||r|r|r|r|r|r|r|r|r|r|r|r|r|r|r|r|r}
& \begin{sideways}\begin{minipage}{1.1cm}
                   Books and Reference
                  \end{minipage}
\end{sideways}& \begin{sideways}Business\end{sideways}& \begin{sideways}Comics\end{sideways}& \begin{sideways}Communication\end{sideways}& \begin{sideways}Education\end{sideways}& \begin{sideways}Entertainment\end{sideways}& \begin{sideways}Family\end{sideways}& \begin{sideways}Finance\end{sideways}& \begin{sideways}\begin{minipage}{1.6cm}Game (Action, Adventure, \\ Arcade, Board)\end{minipage}\end{sideways}& \begin{sideways}Health and Fitnes\end{sideways}& \begin{sideways}Lifestyle\end{sideways}& \begin{sideways}Media and Video\end{sideways}& \begin{sideways}\begin{minipage}{1.6cm}News and\\ Magazines\end{minipage}\end{sideways}& \begin{sideways}Personalisation\end{sideways}& \begin{sideways}Productivity\end{sideways}& \begin{sideways}Tools\end{sideways}& \begin{sideways}Travel and Local\end{sideways}\\\hline\hline
Marginal apps(\%) &43.28 &38.16&31.94&41.2&38.78&38.63&49.81&47.2&38.62&42.54&40.3&46.1&41.35&40.15&44.78&43.97&39.62\\\hline
\textbf{From the rest:}\\
Hot apps(\%) &1.33 &2.65&4&7.44&.81&2.33&4.65&4.4&1.73&3.92&3.17&6.5&4.74&2.25&6.78&5.98&2.95\\
Dominant apps(\%) &.04 &.06&0&.14&0&.03&0&.03&0&0&.07&.11&.07&.02&.06&.07&0\\
Flops(\%) &.25 &.74&.8&1.91&.03&.52&.9&.63&.52&.36&.37&0&.95&.74&2&1.51&.64\\\hline
Mean Eucl. Dist. &.974
&.967
&.914
&.906
&.996
&.958
&.921
&.961
&.938
&.967
&.957
&.947
&.926
&.961
&.926
&.925
&.958
\\
Variance &.010
&.012
&.018
&.027
&.008
&.013
&.020
&.016
&.016
&.013
&.014
&.018
&.023
&.015
&.023
&.024
&.012
\end{tabular}
\end{scriptsize}
\end{table*}

Of the apps associated with one of the four main trend patterns, less than 0.1\% gather a constantly high user base ({\em Dominant} apps). 
Fewer than 1\% are {\em Flops}, i.e., apps that have once gathered a very large user base for a short period of time but have then drastically lost popularity. 
Finally, apps associated to the {\em Hot} category account for about 2-7\% of all relevant apps. 
These apps are quickly rising in popularity. 
It might be trendy apps that will become Flops in the Future or also indicate future trending Dominant apps. 
Considering the Euclidean distance of all individual apps to the mean in their respective GooglePlay category, we observe that the mean and variance are in the order of $0.95$ and $0.2$ for all categories. 
\section{Practical Usage of App Trends}
To demonstrate the practical value of our work, we now consider how trend information can affect app recommendations. 
We have implemented AppJoy~\cite{appjoy-2011} as a representative example of current state-of-the-art app recommenders, and compared the recommendations provided by AppJoy against current trend patterns. AppJoy operates on so-called usage scores, which are constructed by aggregating the following information: (i) time elapsed since the last interaction with an app, (ii) frequency of the user interactions with an app, and (iii) total duration of time the user has interacted with an app. Accordingly, AppJoy bases its recommendations on information sources that correspond to metrics which are used by handset-based mobile analytics tools, such as Google Mobile Analytics and Countly.

To generate recommendations, AppJoy uses a Slope One Prediction model that compares a user's profile to other users with similar application usage history. Formally, we define $S(u)$ as the set of applications used by user $u$. Given an application $i$ and user $u$, We define $R_{u,j}$ as the set of relevant applications $j$ used by other users together with $i$, i.e., $R_{u,j} = \{i | i \in S(u), j \notin S(u), \# S_{i,j} > 0 \}$ where $S_{i,j}$ is the set of users who have used both $i$ and $j$. The relevance of application $j$ for user $u$ is then given by:
\begin{align}
P(u_j) = \frac{1}{size(R_{u,j})} \sum_{i \in R_{u,j}} (dev_{i,j} + u_i).
\end{align}
Here $dev$ is the average of the usage scores between users who have used both $i$ and $j$:
\begin{align}
	dev_{i,j} = \sum_{w \in S_{i,j}} \frac{\upsilon_{w \vdash j} - \upsilon_{w \vdash j}}{size(S_{j,i})}.
\end{align}
Given the relevance scores $P(u_j)$, AppJoy returns the top$-N$ items with the highest score as recommendations.

To illustrate the value of trend and lifecycle information, we ran the AppJoy recommender and our trend analysis for a subset of the data containing $4,500$ users and $1,000$ most frequently used applications in the dataset. As our test period we selected October 2014, due to little seasonal fluctuations, and as training data we selected all data accumulated between January 2014 and September 2014. Given the test data, we used AppJoy to generate recommendations in an incremental fashion for each week.
In particular, we generated recommendations for the first week, then included the data from this period in the training data and generated recommendations for the second week, and so on. We also generated trends for every week, taking into account the lifecycles of the past year, starting from January 1st 2014. 
We counted (i) how many recommended applications can be classified as {\em Flop} or {\em Hot} apps, (ii) how these compare with the total number of {em Flop} and {\em Hot} apps in the top 1000 applications. 
We also calculated temporal diversity, novelty, and accuracy for the recommendation lists~\cite{Lathia:2010}. 
Diversity presents how the recommendations change over time, whereas novelty describes how many new recommendations there are seen compared to the later ones. Novelty of the recommendations relates closely to the trends, because changes in trends should affect new recommendations. Formally these metrics are defined as follows:
\begin{align}
diversity(L_1, L_2, N) &= \frac{|L_2 \setminus L_1|}{N} \\
novelty(L_1, N) &= \frac{|L_1 \setminus A_t|}{N} \\
accuracy(L_1, A) &= \frac{size(L1 \cap A)}{size(A)}
\end{align}

%
\begin{table}
	\caption{Statistics of 20 best recommendations from 1000 applications during a month.}
	\label{tab:recs}
	\centering
	\begin{scriptsize}
		\begin{tabular}{r|r|r|r|r|r|r|r|r|r|r}
			W. & \begin{minipage}{0.4cm}Rec.\\ Stars\end{minipage} & \begin{minipage}{0.4cm}Rec.\\ Flops\end{minipage} & \begin{minipage}{0.4cm}Total \\Stars\end{minipage} & \begin{minipage}{0.4cm}Total\\ Flops\end{minipage} & Div. & Nov. & Acc. & \begin{minipage}{0.4cm}Div. w/o \\flops\end{minipage} & \begin{minipage}{0.4cm}Nov. w/o \\flops\end{minipage} & \begin{minipage}{0.4cm}Acc. w/o \\flops\end{minipage} \\ \hline
			1 & 8  & 5 & 219 & 163 & -	  & - 	 & 0.02 & -    & -    & 0.02 \\
			2 & 7  & 6 & 229 & 158 & 0.80 & 0.98 & 0.03 & 0.90 & 0.90 & 0.12 \\
			3 & 8  & 7 & 232 & 154 & 0.62 & 0.81 & 0 & 0.54 & 0.73 & 0.10 \\
			4 & 10 & 9 & 225 & 150 & 0.56 & 0.75 & 0.11 & 0.50 & 0.68 & 0.11 \\
		\end{tabular}
	\end{scriptsize}
\end{table}
Results of our analysis are shown in Table~\ref{tab:recs} for the top-$20$ recommendations given to all users. 
The results indicate that the number of {\em Hot} apps recommended for each week is quite small and comparable to the number of {\em Flops} recommended in the same time. 
Given that we have generated in total $90,000$ recommendations for $4,500$ users each week, the amount of {\em Hot} recommended corresponds to a very small percentage of the entire set of recommendations. 
In the data set of 1000 apps, more than 200 applications each week can be classified as {\em Hot}, and about 160 applications as {\em Flop}. 
On average, only 3.6\% {\em Hot} apps are recommended, compared to 4.3\% {\em Flops}. 
When {\em Flops} are removed from the recommendations, both novelty and diversity decrease, but accuracy increases slightly. The main reason for this behaviour is that the metrics used by AppJoy to generate recommendations require sufficient amount of usage before an app is recommended. 
However, once sufficient usage has been observed, the app can already be past its "best before" date as the recommendation model does not separate between {\em Hot} and {\em Flop} apps. 
Integrating usage trend information as part of the recommendation process can help to overcome this issue and improve the overall quality of recommendations.



In summary, our analysis clearly indicates that recommendations provided by AppJoy do not reflect dynamics in actual application usage. Other state-of-the-art recommenders, such as Djinn~\cite{Karatzoglou-cikm-2012} and GetJar~\cite{getjar-2012}, are based on similar usage information and are hence likely to exhibit similar patterns compared to usage trends. 
To facilitate users to discover up and coming applications, and to help them avoid apps that are long past their popularity peak, the trend information could be integrated as part of the recommendation process, for instance, by considering it as part of the usage scores used by AppJoy or considering more complex dynamics models, for example, as part of latent factor models~\cite{koren09collaborative}.

\subsection{Application Potential}
We have demonstrated the benefits of considering mobile app trend information for mobile analytics and app recommender systems. 
Another use for trend information is providing developers early feedback about the current popularity of their applications, which they can then use to take countermeasures against negative popularity fluctuations. Trend state can further be correlated with other factors, such as usability gathered through interaction metrics~\cite{ravindranath-osdi-2012}, to provide more detailed feedback of the possible reasons in popularity fluctuations. 


Beyond providing app developers with tools to understand the state of their app, trend can also be used for marketplace analytics to support advertising strategies. On the device side, trend status can be used as an additional metric to identify most redundant applications for removal to reduce clutter on the user interface. The app-filter also enables detecting apps that are rapidly gaining in popularity, which could be used, for instance, by in-app advertisers to entice new app developers as customers or for dynamic pricing models. 

Application trends are also potentially a powerful source of information for characterizing and understanding user interactions, and trend information can be used to support user modelling. 
For example, users with consistently many {\em Hot} applications are continually shifting their application usage, whereas those with many {\em Flop} or {\em Marginal} apps are likely to remain faithful to the apps they originally chose. 


\subsection{Limitations}
Our measurements were collected using a custom mobile application, which itself is naturally prone to retention. 
Carat collects measurements continuously in the background of the mobile device, but only sends the data when launched. 
Accordingly, as long as the user launches the application {\em once} after a sufficiently long period from initial use, we obtain sufficient data to carry out our analysis. To further limit potential biases caused by users stopping to use Carat, we only considered users who had used Carat over a sufficiently long period, e.g. a month.

While comparing the popularity of apps within a category, we relied on category information extracted directly from Google Play. 
On Google Play, the categorization of an app is the responsibility of the developer, and consequently similar apps are likely to contain variations in their categorizations. 
An alternative would be to rely on topic models to derive a categorization of the apps; e.g., Gorla et al.~\cite{Gorla:2014} have demonstrated the use of Latent Dirichlet Allocation (LDA) for mining categories from app store data. Alternatively, trend information could be integrated as part of the topic models together with additional factors, such as number and nature of ratings, and contents of user reviews.

\section{Summary and Conclusion}
We have presented the first ever independent study of retention rates in the wild. 
Our analysis shows that, on average, applications lose $70\%$ of their users in the first week, but the effect is mediated by overall user count as applications with over $1,000$ users show much higher retention rates. 
We also demonstrated that, contrary to reports in the literature, severe losses in usage are rare, with less than $10\%$ of apps losing over $80\%$ of users in the first week. 
We demonstrate that retention rates are an insufficient metric of an application's success as they ignore effects of seasonality and external factors. 
In particular, we demonstrated that applications follow different trend patterns which are not captured by retention. 

As second contribution, we proposed a novel app-filter that can categorize applications according to their currently followed usage trend. 
We focused on four characteristic trends: {\em Marginal} apps with only few users, {\em Dominant} applications of permanent high popularity, {\em Hot} apps with rapidly increasing popularity, and {\em Flop} apps with drastic drop of the usage.
We observed that about 40\% of the apps are {\em Marginal}. 
We analysed application categories from Google Play and show that, for example, during the year 2014, 7.5\% of communication apps have been {\em Hot}, only 0.1\% were {\em Dominant}, and almost 2\% were {\em Flops}. 
This kind of analysis can, in the future, lead application developers to follow needs and desires of the users in faster pace. 

As a practical use case of our work, we considered how our trend-filter can benefit mobile app recommenders by enabling recommendations to focus on those apps that are rising in popularity. 
Trend pattern analysis can be used to strengthen existing heuristics such as interaction rate, download counts, and reviews, and even give more direct way to produce in the wild recommendations taking into account the usage history and trend-pattern of the application. 
Our analysis shows that only $3.6\%$ of the recommendations are for apps which are currently rising in popularity, and that overall recommendations have low novelty and temporal diversity. 
We also demonstrate that the accuracy of the recommendations can be improved by considering trend information. 

\bibliographystyle{abbrv}

\end{document}